\documentclass[11pt]{article}
\usepackage{amsmath, xypic, graphicx}
\usepackage{stmaryrd,xspace,amssymb}
\usepackage{url}

\newcommand{\downset}{\ensuremath{\mathop{\downarrow\!}}}

\newcommand{\Q}{\mathbb{Q}}

\newcommand{\beq}{\begin{equation}}
\newcommand{\eeq}{\end{equation}}
\newcommand{\bea}{\begin{eqnarray}}
\newcommand{\eea}{\end{eqnarray}} \newcommand{\nn}{\nonumber}

\newcommand{\Sets}{\mbox{\textbf{Sets}}}

\newcommand{\ca}{C*-algebra}

\newcommand{\Hs}{Hilbert space}

 \newcommand{\til}{\tilde}
\newcommand{\raw}{\rightarrow}

 \newcommand{\Raw}{\Rightarrow}
 
\newcommand{\lraw}{\leftrightarrow}

 \newcommand{\wed}{\wedge}

\newcommand{\x}{\times}

\newcommand{\inv}{^{-1}}

\newcommand{\er}{\eqref}
 
 \newcommand{\Gm}{\Gamma}
 \newcommand{\Dl}{\Delta}

\newcommand{\lm}{\lambda} 
 \newcommand{\sg}{\sigma}
\newcommand{\Sg}{\Sigma}  \newcommand{\ph}{\phi}
 \newcommand{\phv}{\varphi}
\newcommand{\ch}{\chi} \newcommand{\ps}{\psi} \newcommand{\Ps}{\Psi}
 \newcommand{\Om}{\Omega}
\newcommand{\CA}{{\mathcal A}} \newcommand{\CB}{{\mathcal B}}

   \newcommand{\CL}{{\mathcal L}}

\newcommand{\CO}{{\mathcal O}} \newcommand{\CP}{{\mathcal P}}
 
\newcommand{\CT}{{\mathcal T}} 
 
\newcommand{\C}{{\mathbb C}} 
 
\newcommand{\N}{{\mathbb N}} \newcommand{\R}{{\mathbb R}}
 
\newcommand{\alg}[1]{\ensuremath{#1}}

\newcommand{\id}[1]{\ensuremath{\mathrm{id}}}

\newcommand{\Idl}{\ensuremath{\mathrm{Idl}}}

\newcommand{\context}{\ensuremath{\mathcal{C}}}

\newcommand{\asstopos}{\ensuremath{\mathcal{T}}}

\newcommand{\sa}{\ensuremath{_{\mathrm{sa}}}}

\newcommand{\field}[1]{\ensuremath{\mathbb{#1}}}

\newcommand{\uS}{\underline{\Sigma}}
\newcommand{\uA}{\underline{A}}

\renewcommand{\CA}{\mathcal{C}(A)}
\newcommand{\TA}{\mathcal{T}(A)}
\newcommand{\ie}{\textit{i.e.}}
\newcommand{\eg}{\textit{e.g.}}

\newcommand{\ulA}{\underline{A}}
\renewcommand{\TA}{\asstopos(\alg{A})}
\renewcommand{\CA}{\context(\alg{A})}
\newtheorem{theorem}{Theorem}
\newtheorem{lemma}[theorem]{Lemma}
\newtheorem{proposition}[theorem]{Proposition}

\newtheorem{definition}[theorem]{Definition}

\newenvironment{proof}[1][Proof]%
{ \begin{trivlist}%
  \item[\hskip \labelsep {\bfseries #1}]%
}%
{ \end{trivlist}%
}
\newcommand{\qed}{\nobreak\hfill$\Box$}

\topmargin = - 0.5 cm
\newcommand{\Dom}{\ensuremath{\mathrm{Dom}}}
\newcommand{\slim}{\ensuremath{\mathop{\text{s-lim}}}}

\newcommand{\groter}[2]{\ensuremath{[#1>#2]}}

\long\def\symbolfootnotemark[#1]{\begingroup%
\def\thefootnote{\fnsymbol{footnote}}\footnotemark[#1]\endgroup}

\long\def\symbolfootnotetext[#1]#2{\begingroup%
\def\thefootnote{\fnsymbol{footnote}}\footnotetext[#1]{#2}\endgroup}

\newcommand{\uOm}{\underline{\Omega}}

\newcommand{\ONS}{\mathcal{O}(\Sigma)}
\newcommand{\uSg}{\underline{\Sigma}}

\newcommand{\vNa}{von Neumann algebra}
\newcommand{\CD}{\mathcal{D}}
\newcommand{\Pow}{\operatorname{Pow}}

\newcommand{\eqcomment}[1]{\ensuremath{\tag*{\mbox{\scriptsize{#1}}}}}
\begin{document}
\title{Bohrification of operator algebras and quantum logic}
\author{
Chris Heunen\symbolfootnotemark[1] \symbolfootnotemark[2]
 \and Nicolaas P. Landsman\symbolfootnotemark[1]
 \and Bas Spitters\symbolfootnotemark[3]
}
\symbolfootnotetext[1]{
   Radboud Universiteit Nijmegen,
   Institute for Mathematics, Astrophysics, and Particle Physics,
Heyendaalseweg 135, 6525 AJ    NIJMEGEN, THE NETHERLANDS.
}
\symbolfootnotetext[2]{
   Radboud Universiteit Nijmegen,
   Institute for Computing and Information Sciences, Heyendaalseweg 135, 6525 AJ
NIJMEGEN, THE NETHERLANDS.
}
\symbolfootnotetext[3]{
   Eindhoven University of Technology,
   Department of Mathematics and Computer Science, P.O. Box 513, 5600
   MB EINDHOVEN, THE NETHERLANDS.
}\maketitle
\vspace*{-0.75cm}
\smallskip
\begin{abstract}
Following Birkhoff and von Neumann, quantum logic has traditionally been based on the lattice of closed linear subspaces of some Hilbert space, or, more generally, on the lattice of projections in a von Neumann algebra $A$. Unfortunately, the logical interpretation of these lattices is impaired by their nondistributivity and by various other problems. We show that a possible resolution of these difficulties, suggested by the ideas of Bohr, emerges if instead of single projections one considers elementary propositions to be  families of projections indexed by a partially ordered set $\CA$ of appropriate commutative subalgebras of $A$. In fact, to achieve both maximal generality and ease of use within topos theory, we assume that $A$ is a so-called Rickart C*-algebra and that $\CA$
consists of all unital commutative Rickart C*-subalgebras of $A$.
Such families of projections form a Heyting algebra in a natural way, so that the associated propositional logic
is intuitionistic: distributivity is recovered at the expense of the law of the excluded middle.

Subsequently, generalizing an earlier computation for $n\x n$ matrices, we prove that the Heyting algebra thus associated to $A$ arises as a basis for the internal Gelfand spectrum (in the sense of Banaschewski--Mulvey)  of the ``Bohrification'' $\uA$ of $A$, which is a commutative  Rickart C*-algebra in the topos of functors from $\CA$ to the category of sets.
We explain the relationship of this construction to  partial Boolean algebras and Bruns--Lakser completions.
Finally, we establish a connection between probability measures on the lattice of projections on a Hilbert space $H$
and probability valuations on the internal Gelfand spectrum of $\uA$ for $A=B(H)$.
\end{abstract}

\newpage
\section{Introduction}
As its title is meant to suggest, this paper is an attempt to reconcile the views on the logical structure of quantum mechanics of Niels Bohr on the one hand, and John von Neumann on the other.  This is not an easy task, as indicated, for example,  by the following two quotations:
\begin{quote}
`All departures from common language and ordinary logic are entirely avoided by reserving the word ``phenomenon'' solely for reference to unambiguously communicable information, in the account of which the word ``measurement'' is used in its plain meaning of standardized comparison.' (Bohr \cite{Bohrlogic})
\end{quote}
\begin{quote}
`The object of the present paper is to discover what logical structure one may hope to find in physical theories which, like quantum mechanics, do not conform to classical logic.' (Birkhoff and von Neumann \cite{BvN})
\end{quote}
Another difference lies in the highly technical and advanced mathematical nature of von Neumann's writings on quantum theory, compared with the
philosophical (if not mystical) style of Bohr, who in particular used only very basic mathematics (if any) \cite{Bohr1}. This discrepancy implies that any attempt at  reconciliation between these authors has to rely on mathematical extrapolations of Bohr's ideas that cannot really be justified by his own writings. So be it.

It should be mentioned that in what follows, we use the so-called {\it semantic approach} to the axiomatization of physical theories \cite{Suppe,vF1}, in which theories are defined through their class of models (so that a preceding stage
involving an abstract logical language is lacking).  This, incidentally, is exactly the way quantum mechanics was axiomatized by von Neumann \cite{vN}, who may therefore be seen as a predecessor of the semantic approach (in contrast with Hilbert \cite{Hilbert},
who is regarded as the founder of the  {\it syntactic  approach} to  axiomatization in general).

The outline of this paper is as follows.  The next section reviews
the logic of classical physics from a semantic perspective. We then recall in Section \ref{ss1.2}
how Birkhoff and von Neumann were led to (if not seduced by) their concept of quantum logic, which we criticize and to which we propose an
intuitionistic alternative in Section \ref{sec4}. Von Neumann not only invented quantum logic, he also generalized \Hs\ theory to the theory of operator algebras.
In Section \ref{genOA} we explain the connection between quantum logic and operator algebras, where we take the unusual step of going beyond \vNa s. In fact, we propose to study both traditional quantum logic and  our own intuitionistic version of it in the setting of so-called Rickart C*-algebras. This class of C*-algebras is studied in detail in Sections \ref{RA1} and \ref{RA2}, particularly
with a view on their internalization to topos theory. Specifically, we develop an internal Gelfand theory for commutative
 Rickart C*-algebras, which refines the work of Banaschewski and Mulvey \cite{BM} for general commutative C*-algebras to the Rickart case.
 Section \ref{secor} studies the relationship between our version of intuitionistic quantum logic and partial Boolean algebras
 on the one hand, and so-called Bruns--Lakser completions on the other. Finally, in
 Section \ref{finalsection}
 we explain how the well-known concept of a probability measure on the projection lattice on a \Hs\ is related to various
 concepts intrinsic to our approach, and explicitly compute a non-probabilistic state-proposition pairing.

This paper is a continuation of our earlier work \cite{HLS,caspersheunenlandsmanspitters:nlevelsystem}, which provides
some background, particularly on quantum theory in a topos. However, the present paper is largely self-contained
and takes our program a significant step further.
\section{The logic of classical physics}\label{ss1.1}
To explain the basic issue, we first recall the logical structure of classical physics.\footnote{It is remarkable that this structure was not written down by either Boole or Hamilton in the mid 19th Century, as it clearly emerges from the conjunction of their ideas on propositional logic and on classical physics, respectively \cite{Boole,Hamilton}. As far as we know, however,  the logical structure of classical physics was first explicated by Birkhoff and von Neumann in 1936  \cite{BvN}; see also \cite{Redei} for a very clear account.}
Let $X$ be the phase space of a classical physical system; we assume that $X$ is a topological space with ensuing Borel  structure.  We identify elements of $X$ with (pure) states of the system. Observables are measurable functions $f:X\raw\R$, and elementary propositions take the form $f\in\Dl$, where $\Dl$ is a measurable subset of $\R$. Further propositions
are inductively built from these through the operations $\neg$ of negation, $\vee$ of disjunction and $\wed$ of conjunction.
An elementary  proposition $f\in\Dl$
is dictated by physics to be true in a state $x\in X$ iff $f(x)\in \Dl$, i.e.\ iff $x\in f\inv(\Dl)$; this notion of truth is defined semantically (as opposed to formal derivability in the syntactic approach). Consequently, we may
introduce the notation $\models$ of semantic entailment, meaning (sic) that
$(f\in\Dl)\models (g\in\Gm)$ whenever  the truth of $f\in\Dl$ implies the truth of $g\in\Gm$.
Hence one may form the associated Lindenbaum--Tarski algebra of equivalence classes $[f\in \Dl]$, where we say that $(f\in \Dl) \sim (g\in \Gm)$ when $(f\in\Dl)\models (g\in\Gm)$ and $(g\in\Gm) \models (f\in\Dl)$ both hold (in words, $f\in\Dl$ is true
iff $g\in\Gm$ is true). This yields the identification $[f\in
\Dl]\cong  f\inv(\Dl)$ and the ensuing identification of the
Lindenbaum--Tarski algebra of the given system with the Boolean
algebra\footnote{Recall that a lattice $L$ is called
{\it orthocomplemented} when there exists a map $\perp \colon L \to L$ that satisfies
$x^{\perp\perp} = x$, $y^{\perp} \leqslant x^\perp$ when $x \leqslant
y$,  $x \wedge x^\perp = 0$, and $x \vee x^\perp = 1$. For example,
the lattice of closed subspaces of a Hilbert space has an
orthocomplement; namely, $V^\perp$ is the orthogonal complement of $V$.
A lattice $L$ is called
{\it Boolean} when it is distributive and orthocomplemented, in which
case the orthocomplement $\perp$ is called a {\it complement} and
written as $\neg$, and has the logical meaning of negation.}
 $\Sg(X)$ of (Borel) measurable subsets of $X$. Under this identification, the logical connectives
$\models$,
$\neg$, $\vee$ and $\wed$ descend to set-theoretic inclusion $\subseteq$, complementation $(-)^c$, union $\cup$,  and intersection $\cap$, respectively, and these are compatible in that $\cup$  and $\cap$ are precisely the lattice operations
sup and inf induced by the partial order $\subseteq$. Finally, $\Sg(X)$ has bottom and top elements $\emptyset$ and
$X$, respectively, which play the role of falsehood $\bot$ and truth $\top$, and with respect to which $(-)^c$ is an orthocomplementation.
This means, in particular, that besides the law of contradiction $p\wed(\neg p)=\bot$, which in this case descends to
$p\cap p^c=\emptyset$, one has the law of excluded middle $p\vee(\neg p)=\top$,  descending to $p\cup p^c=X$.

This procedure is unobjectionable, in that $\neg$, $\vee$ and $\wed$ thus interpreted in set theory indeed have their usual meaning
of negation,  disjunction, and conjunction, respectively. In particular (identifying propositions with their image in $\Sg(X)$),
\begin{enumerate}
\item  Disjunction and conjunction distribute over each other;\footnote{I.e., $p\wed (q\vee r)=(p\wed q)\vee (p\wed r)$ and
$p\vee (q\wed r)=(p\vee q)\wed (p\vee r)$.}
\item $p\vee q$ is true iff $p$ is true or $q$ is true;
\item $p\wed q$ is true iff $p$ is true and $q$ is true;
\item $\neg p$ is true iff $p$ is not true;
\item There is a material  implication $\Raw: \Sg(X)\times\Sg(X)\raw\Sg(X)$ that
satisfies\footnote{If $\Sg(X)$ is seen as a category (with a unique arrow from $p$ to $q$ iff $p\leq q$), then $\Raw$ is right adjoint to $\wed$.\label{rafoot}}
\beq p\leqslant (q\Raw r)\:\:\mbox{iff}\:\: p\wed q\leqslant r,\label{mi} \eeq
namely
$(q\Raw r)=(q^c\cup r)$.
\end{enumerate}
\section{The lure of quantum logic}\label{ss1.2}
The quantum logic of  Birkhoff and von Neumann~\cite{BvN} is an attempt to adapt this scheme to
quantum mechanics.\footnote{See, for example, \cite{dallachiaragiuntini:quantumlogics,DC2,Redei} for
 recent surveys of quantum logic in the tradition of  Birkhoff and von Neumann. The relationship
between quantum logic and projective geometry, which was a major discovery of von Neumann's, is
beautifully surveyed in \cite{SvS}. A good philosophical critique of quantum logic is
\cite{Stairs}.}
This time, the starting point is a Hilbert space $H$, whose unit vectors $\Psi$ are interpreted as pure states. Furthermore,
observables are taken to be self-adjoint operators $a:\Dom(a)\raw H$, with dense domain $\Dom(a)\subseteq H$; in what follows, we assume for simplicity that $\Dom(a)=H$, so that $a$ is bounded.
Elementary propositions assume the same form ``$a\in\Dl$'' as in classical physics, and may formally be combined using the connectives $\neg$ , $\vee$, and $\wed$. This time, the  truth predicate on  $a\in\Dl$ is determined by the associated spectral projection, which we write as $E_a(\Dl)$ (so that the map $\Dl\mapsto E_a(\Dl)$ is the spectral measure defined by $a$).
According to von Neumann \cite{vN}, the proposition $a\in\Dl$ is true in a state $\Psi\in H$ iff $\Psi\in E_a(\Dl)H$, so that the equivalence classes determined by this truth condition may be written as $[a\in\Dl]=E_a(\Dl)H$. Each such class is a closed linear subspace of $H$, and semantic entailment of propositions obviously descends to inclusion of closed linear subspaces.
Thus it is hard to resist the temptation to conclude that the lattice $\CL(H)$ of closed linear subspaces of the Hilbert space $H$ (with partial ordering given by inclusion) is the correct quantum-mechanical analogue of the lattice $\Sg(X)$ of measurable subsets of the classical phase space $X$.

Birkhoff and von Neumann \cite{BvN} were indeed seduced by this perspective, and proposed  that the logic of quantum mechanics is described by the lattice structure of $\CL(H)$, which, then, plays the role of the  Lindenbaum--Tarski algebra of equivalence classes of quantum-mechanical propositions \cite{Redei}. Once more using the same notation for the images of propositions and logical connectives in $\CL(H)$ as for these things themselves, the ensuing lattice operations on $\CL(H)$ are given by
$p\vee q=p\dot{+}q$ (i.e.\ the closure of the linear span of $p$ and $q$) and $p\wed q=p\cap q$. As to negation,
Birkhoff and von Neumann decided to define $\neg p$ as the proposition that is true whenever $p$ is false; unlike in classical physics, this is not the same as saying that $p$ is not true. Now in quantum mechanics a proposition $a\in\Dl$ is false in a state $\Psi$ iff $\Psi\in (E_a(\Dl)H)^{\perp}$ (where $(-)^{\perp}$ denotes the orthogonal complement), so that $\neg p=p^{\perp}$. With the bottom and top elements of $\CL(H)$ given by $\{0\}$ and $H$, respectively, this implies that
$\neg$ is an orthocomplementation, so that
the quantum logic of \cite{BvN} formally satisfies both the law of contradiction,
implemented as
$p\cap p^{\perp}=\{0\}$, and the law of excluded middle $p \dot{+} p^{\perp}=H$.

Nonetheless, we feel that Birkhoff and von Neumann should have resisted this temptation.\footnote{In what follows, we intend to criticize the {\it logical} interpretation of the connectives $\vee,\,\wed,\,\neg$ in standard quantum logic; we do not take issue with their {\it operational} interpretation assigned by the Geneva school led by Piron \cite{moore,Piron}.}
 Indeed, compared with the five points in favour of the propositional logic of classical physics being the  Boolean algebra of  measurable subsets of phase space, we now have:
\begin{enumerate}
\item  Disjunction and conjunction do not distribute over each other;\footnote{The lattice $\CL(H)$ does satisfy a weakening of distributivity called \emph{orthomodularity}; see Section \ref{secor}.}
\item There are states in which $p\vee q$ is true while neither $p$ nor $q$ is true;\footnote{Take any unit vector that lies in the subspace spanned by $p$ and $q$ without lying in either $p$ or $q$. This is famously the kind of state Schr\"{o}dinger's Cat is in. }
\item There are propositions $p$ and $q$  for which  $p\wed q$ cannot be regarded as the conjunction of $p$ and $q$ because this conjunction is physically undefined;\footnote{Take, for example, $q$ to be a spectral projection for position and $p$ to be one for momentum, or, more generally, any pair of projections that do not commute.}
\item $\neg p$ is true iff $p$ is false, rather than iff $p$ is not true;\footnote{The distinction between ``false'' and ``not true'' arises from the Born rule of quantum theory, according to which  the proposition $a\in\Dl$ is true in a state $\Psi\in H$
with probability  $\| E_a(\Dl)\Psi\|^2$. If this probability equals one we say the proposition is true, and if it equals zero we say it is false. Hence ``not true'' refers to all probabilities in the semi-open interval $[0,1)$, rather than to zero alone.}
\item There exists no map  $\Raw: \CL(H)\raw\CL(H)$ that  satisfies  \er{mi}.
\end{enumerate}
It is important to realize that the equality  $p\vee(\neg p)=\top$  is only true in quantum logic because neither $\vee$ nor $\neg$ has its usual logical meaning. In fact,  in quantum logic this equality only formally expresses the law of excluded middle; it   is semantically empty.

As to the last point, it can be shown that one has a material implication on an orthocomplemented lattice $\CL$ (i.e.\ a map  $\raw: \CL\raw\CL$ satisfying \er{mi}) iff $\CL$ is Boolean, in which case
$p\Raw q=\neg p\vee q$;  see, e.g.,  \cite[Prop.\ 8.1]{Redei}. Consequently, quantum logicians tend to weaken the property \er{mi} by requiring it only for all $q$ and $r$ that are compatible in the sense that $q = (q \wedge r^\perp)
\vee (q \wedge r)$; in $\CL(H)$ this is the case iff $q$ and $r$ commute. If $\CL$ is orthocomplemented, the existence of such an implication forces $\CL$ to be orthomodular and implies that $\Raw$ takes the form of the ``Sasaki hook''
\beq  p \Raw_S q = p^\perp \vee (p \wedge q),\label{hook}
\eeq
discussed in some detail in Section \ref{secor} below.

In order to pave the way for the algebraic ideas to follow, we close this section by reminding the reader of the well-known connection between closed linear subspaces of $H$ and  projections  $p$ on $H$, defined as  bounded linear operators $p:H\raw H$ satisfying $p^2=p^*=p$. Indeed, we know from elementary Hilbert space theory that there is a bijective  correspondence between projections $p$ on $H$ and closed linear subspaces of $H$: a projection $p$ defines
such a subspace as its image $pH$, and any closed linear subspace is the image of a unique projection.
For consistency with later notation, we denote the set of  all projections on $H$ by $\CP(B(H))$ (instead of the more natural expression $\CP(H)$), where $B(H)$ is the algebra of all bounded operators on $H$.
 If we now define a partial order on the set $\CP(B(H))$ of
 $p\leq q$ iff $pH\subseteq qH$, by construction we obtain a lattice isomorphism
\beq \CP(B(H))\cong \CL(H). \label{latiso}\eeq
In view of this, if no confusion can arise we make no notational distinction between closed linear subspaces and projections, denoting both by $p$ etc.
The partial order on $\CP(B(H))$ may, in fact be defined without reference to \er{latiso}: one has
\beq p\leq q \mbox{ iff }  pq=qp=p. \label{leq}\eeq
As to the ensuing lattice operations, defining
\beq p^{\perp}=1-p,\label{perp}\eeq
the inf and sup derived from $\leq$ may be expressed by
\begin{eqnarray}
p\wed q&=&\slim_{n\raw\infty} (pq)^n; \label{inf}\\
p\vee q&=&(p^{\perp}\wed q^{\perp})^{\perp}, \label{sup}
\end{eqnarray}
where $\slim$ denotes the limit in the strong operator topology.\footnote{The strong operator topology on $B(H)$ is induced by the seminorms $p_{\Psi}(a)=\| a\Psi\|$, $\Psi\in H$, so that $\slim_n a_n=a$ iff $\lim_n \| (a_n-a)\Psi\|=0$
for all $\Psi\in H$.} If $p$ and $q$ happen to commute, these expressions reduce to
\begin{eqnarray}
p\wed q&=& pq; \label{infc}\\
p\vee q&=& p+q-pq. \label{supc}
\end{eqnarray}
\section{Intuitionistic quantum logic}\label{sec4}
We now return to Bohr for guidance towards the solution
of the problems with von Neumann's quantum logic.
Bohr's best-known formulation of what came to be called his ``doctrine of classical concepts'' \cite{Scheibe} is as follows:
\begin{quote}
`However far the phenomena transcend the scope of classical physical explanation, the account of all evidence must be expressed in classical terms. (\ldots) The argument is simply that by the word {\it experiment} we refer to a situation where we can tell others what we have done and what we have learned and that, therefore, the account of the experimental arrangements and of the results of the observations must be expressed in unambiguous language with suitable application of the terminology of classical physics.' \cite{bohr49}
\end{quote}

For simplicity, we assume in this section that our Hilbert space $H$ is $n$-dimensional with $n<\infty$;
the general case will be covered in the remainder of the paper. Anticipating later generalizations at least in the notation, we  write $A=M_n(\C)$  for the algebra of $n\x n$ matrices.
Our mathematical translation of Bohr's doctrine, then, is to  study $A$  through its {\it
commutative}  subalgebras $C$, where for technical reasons we assume $C$ to contain the unit matrix
and to be closed under the involution $*$
(i.e.\ Hermitian conjugation, often denoted by a dagger by physicists); that is, if $a\in C$, then
$a^*\in C$.
Thus  we define $\CA$ to be
the set  of all unital commutative $\mbox{}^*$-subalgebras of $A$. This set is partially ordered by inclusion, i.e., for $C,D\in\CA$ we say that $C\leqslant D$ iff $C\subseteq D$.  The poset $\CA$ is merely a so-called meet-semilattice rather than a lattice: although infima exist in the form $C\wed D=C\cap D$, there are no suprema, since $C$ and $D$ will not, in general, be contained in a commutative subalgebra of $A$ (unless $cd=dc$ for all $c\in C$ and $d\in D$).

It is much harder to make mathematical sense of Bohr's idea of ``complementarity'', especially as his formulation of this notion remained vague and in fact changed over time.\footnote{The literature on complementarity is abundant, but we recommend the critical studies \cite{Held,Lahti}.} Be it as it may, we interpret the idea of complementarity in the following way:
rather than following von Neumann \cite{vN} in defining an elementary quantum-mechanical proposition as a {\it single} projection on $H$, we follow (the spirit of) Bohr in defining such a proposition as a {\it family}  $\{p_C\}_{C\in\CA}$ of projections, one for each ``classical context'' $C$, with $p_C$ pertinent to that context in requiring that
 $p_C\in \CP(C)$. For the moment, we simply postulate this idea, but in the main body of the paper we will actually derive it from the doctrine of classical concepts (rephrased mathematically as explained above). Adding minimal mathematical structure, our proposal means that  we replace the lattice $\CP(A)$ of all projections in $A$ as the codification of quantum logic by
\beq
\CO(\Sg)=\{S:\mathcal{C}(A) \raw \CP(A)\mid {S}(C)\in \CP(C),\,
{S}(D)\leq {S}(E)\:\mbox{if}\: D\subseteq E\}, \label{bohr}
\eeq
where $\CP(C)$ is the (Boolean) lattice of projections in $C$.
As already mentioned, we regard each $S\in  \CO(\Sg)$ as a single proposition as far as logical structure is concerned; physically, $S$ breaks down into a family $\{S(C)\}_{C\in\CA}$. This could either mean that one invents a question for each context $C$ separately (compatible with the monotonicity in \er{bohr}), or that one constructs such a family from a single proposition in the sense of von Neumann. The latter may be done in at least two ways:
\begin{enumerate}
\item For $p\in\CP(A)$, one defines
\begin{eqnarray}
S_p(C) &=& p \mbox{ if } p\in C;\nn \\
&=& 0 \mbox{ if } p\notin C. \label{ronnie}
\end{eqnarray}
\item One uses the ``inner Daseinisation'' map of D\"{o}ring and Isham \cite{DI}, which associates
the best approximation in each $C$ to a proposition $a\in\Dl$; see also \cite{HLS}. In fact, \er{ronnie} may be seen as a crude analogue of this procedure.
\end{enumerate}

In order to unravel its logical structure, we turn  $\CO(\Sg)$ into
 a poset under pointwise partial ordering with respect to
the usual ordering of projections, i.e.\   for $S,T\in \CO(\Sg)$ we put $S\leqslant T$ iff $S(C)\leq T(C)$ for all $C\in \CA$,
where  $\leq$ is defined by \er{leq}. The main observation is that
$\CO(\Sg)$ is a {\it complete Heyting algebra}\footnote{A  {\it Heyting algebra}
is just a lattice $\CL$ with a map $\Raw:\CL\x\CL\raw\CL$ satisfying
\er{mi}; it is automatically a distributive lattice. It is complete
when $\CL$ is complete as a lattice. The interpretation of $\Raw$ as a right adjoint to $\wed$, as in footnote \ref{rafoot}, remains valid.
 In particular, every Boolean
lattice is a Heyting algebra with $x \Rightarrow y=\neg x \vee y$.}
under this partial ordering. 

The whole point now is that  in being a (complete) Heyting algebra, $\CO(\Sg)$
 defines an {\it intuitionistic propositional logic}, which in fact is not Boolean    \cite{caspersheunenlandsmanspitters:nlevelsystem}.\footnote{
 A Heyting algebra is Boolean iff the negation $\neg$ defined by \er{defneg} below is an orthocomplementation.}
 First, the inf and sup derived from $\leqslant$ are given by the pointwise expressions
\begin{eqnarray}
(S \wed T)(C) &=& S(C) \wed T(C); \\
(S \vee T)(C) &=& S(C) \vee T(C).
\end{eqnarray}
The top and bottom elements are $\top: C\mapsto 1$ and $\bot:C\mapsto 0$ for all $C$, where 1 and 0 are seen as elements of $\CP(C)$.
Material implication  is defined by
\beq
S\Raw T=\bigvee\{U\in \CO(\Sg)\mid U\wed S\leqslant T\}, \label{fHA}
\eeq
and is explicitly given by the nonlocal formula
\beq
(S\Raw T)(C)=\bigwedge^{\CP(C)}_{D\supseteq C}S(D)^{\perp}\vee T(D).
\label{HI}
\eeq
Here the right-hand side
denotes the  greatest lower bound of all $S(D)^{\perp}\vee T(D)$, $D\supseteq C$, that lies in $\CP(C)$.
The derived operation of
negation, which in any Heyting algebra is given in terms of $\Raw$ by
\beq \neg x=(x\Raw\perp), \label{defneg}
\eeq is then equal to
\beq
(\neg S)(C)=\bigwedge^{\CP(C)}_{D\supseteq C}S(D)^{\perp}. \label{Hneg}
\eeq

The natural semantics for the intuitionistic propositional logic $\ONS$ is of Kripke type \cite{Kripke} (see also \cite{Dummett,Gold}). First, we take the Kripke frame to be the poset $\CA$, and denote the set of upper sets in $\CA$ by $\CO_A(\CA)$.\footnote{This notation reflects the fact that the upper sets in a poset just form its Alexandrov topology.}
Each unit vector $\Ps\in\C^n$  defines a state on $A$, i.e.\ a linear functional $\ps:A\raw \C$
that  satisfies $\ps(1)=1$ and $\ps(a^*a)\geq 0$ for all $a\in A$  by $\ps(a)=(\Ps,a\Ps)$; more generally, each density matrix defines a state on $A$ by taking expectation values.
This, in turn, defines a map
\beq
V_{\ps}:\ONS\raw \CO_A(\CA)\eeq
 by\footnote{Note that \er{result} indeed defines an upper set in $\CA$. If $C\subseteq D$ then
${S}(C)\leq {S}(D)$, so that $\ps({S}(C))\leq \psi({S}(D))$ by positivity of
states, so that $\ps({S}(D))=1$ whenever $\ps({S}(C))=1$ (given that $\ps({S}(D))\leq 1$, since $\ps(p)\leq 1$ for  any projection $p$).}
\beq V_{\ps}(S)=\{C\in\CA\mid \ps({S}(C))=1\}.\label{result}\eeq
This map is to be compared with the traditional truth attribution
\beq
W_{\ps}: \CP(A)\raw\{0,1\}\eeq
in quantum logic, given by
$W_{\ps}(p)=1$ iff $\ps(p)=1$.\footnote{This is a slight generalization from the example $A=B(H)$, where
a proposition $p$ is called true in a pure state $\Ps$ if $\Ps\in pH$. This is equivalent to $\ps(p)=(\Ps,p\Ps)=1$.}
Consequently, \er{result}
lists the ``possible worlds'' $C$ in which $S(C)$ is true in the usual sense.

However, unless $A$ is Abelian,  neither $V_{\ps}$ nor $W_{\ps}$ is a lattice homomorphism;\footnote{More precisely, $V_{\ps}$ is not a frame homomorphism, see below.}
even the restrictions of  $W_{\ps}$ to Boolean sublattices of $\CP(A)$ fail to be lattice homomorphisms. In fact, for $n>2$  there are no lattice homomorphisms $W:\CP(A)\raw\{0,1\}$ or $V:\ONS\raw \CO_A(\CA)$ altogether;
the first claim  is the content of the original Kochen--Specker
Theorem \cite{KS},  and the second is its generalization by the
authors \cite{HLS,caspersheunenlandsmanspitters:nlevelsystem} (see
also \cite{DKS,IB} for predecessors of this
generalization).

In any case, we are now in a position to compare the quantum logic of Birkhoff and von Neumann with our own version, at least as far as the five points listed in both Sections \ref{ss1.1} and \ref{ss1.2} are concerned:
\begin{enumerate}
\item The lattice $\ONS$ is distributive;
\item Defining a proposition $S\in\ONS$ to be true in a state $\ps$ if $V_{\ps}(S)=\CA$ (i.e.\ the top element of the Kripke frame $\CO_A(\CA)$), it follows that $S\vee T$ is true iff either $S$ or $T$ is true;\footnote{This has the rather trivial origin that
$V_{\ps}(S)=\CA$ iff $S(\C\cdot 1)=1$, which forces $S(C)=1$ for all $C$.}
\item The conjunction $S\wed T$ is always defined physically, as it only involves ``local'' conjunctions $S(C)\wed T(C)$
for which $S(C)$ and $T(C)$ both lie in $P(C)$ and hence commute;
\item  Defining  $S\in\ONS$ to be false in $\ps$ if $V_{\ps}(S)=\emptyset$ (i.e.\ the bottom element of $\CO_A(\CA)$),
one has that $\neg S$  is true iff $S$ is false.
\item There exists a map  $\Raw: \ONS\raw\ONS$ that  satisfies  \er{mi}, namely \er{HI}.\footnote{Note that, compared with the Sasaki hook \er{hook}, one has $(S\Raw T)(C)\neq S(C)\Raw_S T(C)=S(C)^{\perp}\vee T(C)$, as the left-hand side is nonlocal in $C$.}
\end{enumerate}

To restore the balance a little, let us draw attention to a good side of traditional quantum logic, namely its essentially topological character.
This is especially clear in its original incarnation, where
propositions are identified with closed subspaces of Hilbert
space. This aspect is somewhat obscured in the reformulation in terms
of projections, and looks truly remote in our version
\er{bohr}. However, the lattice defined by \er{bohr} is topological in
a more subtle sense, in that it defines the ``topology'' of a
``pointless space''. To explain this, we note that the topology
$\CO(X)$ on a space $X$ has the structure of a so-called {\it
  frame}\footnote{This notion is not to be confused with that of a
  Kripke frame; the latter is not an instance of the former at all.}, i.e.\ a complete distributive lattice such that
$x\wedge \bigvee_{\lambda}y_{\lambda}=\bigvee_{\lambda}x\wedge
y_{\lambda}$ for arbitrary families $\{y_{\lambda}\}$. Here the partial order on the opens in $X$ is simply given by inclusion.
For a large class of spaces (namely, the so-called sober ones), one may recover $X$ from its frame of opens in two steps:
first, the points of $X$ correspond to the set $\mathrm{pt}(\CO(X))$ of lattice homomorphisms $\phv: \CO(X)\raw \{0,1\}$  that preserve arbitrary  suprema,
and second, the topology is recovered in stating that the open sets in $\mathrm{pt}(\CO(X))$
are those of the form $\{\phv\in \mathrm{pt}(\CO(X))\mid \phv(U)=1\}$, for each $U\in\CO(X)$. Compare the discussion following Proposition \ref{prop:vna}.

Our notation $ \CO(\Sg)$ for the lattice defined by \er{bohr} is meant to suggest that it is a frame, and indeed it is: the Heyting algebra structure of $\ONS$ is actually derived from its frame structure by \er{mi}. More generally, any frame is at the same time a complete Heyting algebra with implication  \er{mi}, and in fact frames and complete
Heyting algebras are essentially the same things.\footnote{
The infinite distributivity law in a frame is
automatically satisfied in a Heyting algebra. Frames and  Heyting algebras do not form isomorphic or even equivalent categories, though, for frame maps do not necessarily preserve the implication $\Raw$ defining the Heyting algebra structure.}
Due to the Kochen--Specker Theorem of \cite{HLS,caspersheunenlandsmanspitters:nlevelsystem}, the frame $\ONS$ cannot be of the type given by the opens of some genuine topological space $\Sg$, but even though it isn't,
one may reason about $\ONS$  {\it as if it were} the collection of opens of a space. This underlying space, $\Sg$, is so to speak ``virtual'', or ``pointfree''; it  only exists through its associated frame $\ONS$.
The upshot is that while a classical physical system has an actual topological space associated with it, namely
its phase space, a quantum system still defines a space, albeit a pointfree one that only exists through its ``topology'', namely the frame defined by \er{bohr}.

Our proposal, then, is that quantum logic should not be described by an orthomodular lattice of the type $\CP(A)$, but by a frame or Heyting algebra of the type \er{bohr}. Thus the ``Bohrification'' of quantum logic is
intuitionistic. In this light, it is interesting to note that Birkhoff and von Neumann actually considered this possibility, but rejected it:
\begin{quote}
`The models for propositional calculi which have been considered in the preceding sections are also interesting from the standpoint of pure logic. Their nature is determined by quasi-physical and technical reasoning, different from the introspective and philosophical considerations which have had to guide logicians hitherto. Hence it is interesting to compare the modifications which they introduce into Boolean algebra, with those which logicians on ``intuitionist'' and related grounds have tried introducing.

The main difference seems to be that whereas logicians have usually assumed that properties [\ldots] of negation were the ones least able to withstand a critical analysis, the study of mechanics points to the {\it distributive identities} [\ldots] as the weakest link in the algebra of logic.' \cite{BvN}
\end{quote}
\section{Generalization to operator algebras}\label{genOA}
The technical thrust of this paper lies in the generalization of the above ideas to infinite-dimensional Hilbert spaces $H$ and to more general algebras of operators than $A=B(H)$. As we shall see, this generalization is  quite interesting mathematically, but we also envisage future physical applications to  infinite quantum systems and other systems with so-called superselection rules \cite{Haag}, as well as to quantization and the classical limit of quantum mechanics \cite{handbook}.

The natural setting for our work is  the theory of operator algebras, created by none other than von Neumann.
 The class of operator algebras he introduced is now aptly called \vNa s (older names are rings of operators and $W^*$-algebras), and  incorporates not only the highly noncommutative world of the $n\x n$ matrices and their infinite-dimensional generalization $B(H)$,  but also covers the commutative case, with a direct link to Boolean algebras and hence classical logic.
The main reference for the general theory of \vNa s is Takesaki's three-volume treatise \cite{Tak1,Tak2,Tak3}; the relationship between \vNa s and quantum logic has been  beautifully described by R\'{e}dei \cite{Redei}.
\begin{definition}
For any Hilbert space $H$, a \emph{von Neumann algebra} of operators on $H$ is a subalgebra $A$ of $B(H)$ that contains
the unit of $B(H)$, contains the adjoint $a^*$ whenever it contains $a$, and in addition satisfies one (and hence both) of the following
equivalent conditions:
\begin{enumerate}
\item  $A''=A$;
\item $A$ is closed in the strong operator topology.\footnote{Here $A\subset B(H)$ is strongly closed if for any strongly convergent net $(a_{\lm})$ in $A$
with limit $a$ in $B(H)$
(in the sense that $\| a_{\lm}\Psi - a\Psi\|\raw 0$ for all $\Psi\in H$),  the limit $a$ in fact lies in $A$. }
\end{enumerate}
\end{definition}
In the first condition, we write $A''\equiv (A')'$, where $A'$ is the commutant of $A$, consisting of all $a\in B(H)$ that commute with any $b\in A$.

To see how \vNa s lead to a generalization of quantum logic \cite{Redei}, we note that a von Neumann algebra is generated by its projections: if
\beq \CP(A)=\{p\in A\mid p^2=p^*=p\}\label{PA}\eeq is the set of projections in $A$, then
$\CP(A)''=A$; equivalently, the strong closure of the (algebraic) linear span of $\CP(A)$ equals $A$.\footnote{
Another good way of looking at \vNa s is to see them as symmetries: any \vNa\ on a Hilbert space $H$ arises as the algebra of invariants of some group action on $H$, in the sense that $A=U(G)'$ for some group $G$ acting on $H$  through a unitary representation $U$. To see this, note in one direction that $U(G)'''=U(G)'$, so that $U(G)'$ is indeed a \vNa. In the opposite direction, given $A$, let
$G$ be the group of all unitary operators in $A'$ and take $U$ to be the defining representation.}
Moreover, for any \vNa\ $A$,  the set $\CP(A)$ is an  orthomodular lattice  under the ordering defined by \er{leq}, with orthocomplementation, inf and sup given by \er{perp}, \er{inf}, and \er{sup}, respectively, and bottom and top elements $\bot=0$, $\top=1$. One may continue to identify $p\in\CP(A)$ with an elementary quantum-mechanical proposition, and look at  $\CP(A)$ as a generalized quantum logic in the sense of Birkhoff and von Neumann. It is important to note that the lattice
$\CP(A)$ is always complete (in that infima and suprema of arbitrary subsets exist).

Inspired by both von Neumann's operator algebras and the theory of commutative Banach algebras, Gelfand and Naimark introduced the concept of a \emph{\ca}\ in 1943. Unlike a \vNa, a C*-algebra is defined without reference to a Hilbert space, namely as an involutive Banach algebra $A$ for which $\| a^*a\|=\|a\|^2$ for each $a\in A$. For any Hilbert space $H$, the algebra $B(H)$ satisfies these axioms. More generally, each \vNa\ is a \ca, but even if
a C*-algebra is concretely given as an algebra of operators on some \Hs, it need not be strongly closed and hence need not be a \vNa.  In fact,
the class of all C*-algebras is not directly relevant to quantum logic, as a generic C*-algebra may not have enough projections.

One can already see this in the commutative case, where (in the unital case) one always has the so-called Gelfand isomorphism 
\begin{equation}
A\cong C(\Sg_A)\equiv C(\Sg_A,\C), \label{Geliso}
\end{equation}
for some compact Hausdorff space $\Sg_A$, called the (Gelfand) {\it spectrum} of $A$. Now, under this isomorphism the projections in $A$ correspond to characteristic functions of (Borel) subsets of $\Sg_A$, so we immediately see that
 if $\Sigma_A$ is connected, $A\cong C(\Sigma_A)$ has no nontrivial projections (i.e., except 0 and 1). 

 For later use, we briefly recall how the isomorphism \er{Geliso} comes about. One may define
$\Sg_A$ as the space  of {\it characters} of $A$, i.e.\  nonzero multiplicative linear functionals  $\phv:A\raw\C$ that satisfy $\phv(ab)=\phv(a)\phv(b)$; such functionals are automatically continuous and hence $\Sg_A$ inherits the weak$\mbox{}^*$-topology on the Banach space dual $A^*$.\footnote{This is the weakest topology under which each  $\hat{a}$ defined below is continuous.}  Subsequently, one defines a map
\begin{eqnarray}
A&\stackrel{\cong}{\raw}& C(\Sg_A); \nn \\
a&\mapsto& \hat{a}; \nn \\
\hat{a}(\phv)&=&\phv(a).
\end{eqnarray}
This map is called the {\it Gelfand transform} and turns out to be an isomorphism when $A$ is a commutative C*-algebra with unit, and $C(\Sg_A)$ is equipped with pointwise operations and the supremum norm.
The space $\Sg_A$ is homeomorphic to the set of all regular maximal ideals of $A$,\footnote{In this context, an ideal $I$ of a commutative Banach algebra $A$ is by definition closed, and is called regular if the quotient algebra $A/I$ admits an identity.}
topologized by letting each $a\in A$ define a basic open that consists of all regular maximal ideals of $A$ not containing $a$.
 The pertinent homeomorphism is then given by $\phv\lraw\phv\inv(\{0\})$.
 
  Interestingly, it is also possible to directly describe
 this topology $\CO(\Sg_A)$ as a frame (up to frame isomorphism), without taking recourse to the initial construction of $\Sg_A$ as a set. In the special case that $A$ has sufficiently many projections, for example, when it is a commutative von Neumann 
 algebra (or, more generally, a commutative Rickart C*-algebra, as in Definition \ref{defRCA} below), 
 this description is given by\footnote{See  \cite{Coq,coquandspitters05,HLS} for the case of general commutative C*-algebras.}
  \begin{equation}
\CO(\Sigma_A)\cong \Idl(\CP(A)),\label{inzicht}
\end{equation}
where $\mathrm{Idl}(L)$ is the usual frame of ideals of a lattice $L$,\footnote{This is the
collection of nonempty lower closed subsets $I\subset  L$ such that $x,y\in I$ implies $x\vee y\in
I$, ordered by inclusion \cite[p.59]{johnstone82}.}  and $\CP(A)$ is the lattice of projections in
$A$, as above (in the present case, where $A$ is assumed to be commutative, this lattice is Boolean,
see below). This result (which may be unfamiliar even to specialists in C*-algebras) is a special
case of Theorem \ref{RCAS} below. 

The absence of sufficiently many projections in a general C*-algebra
inspires the search for extra conditions on a C*-algebra that do have an
ample supply of projections and hence provide a good home for quantum
logic. As we have seen, \vNa s indeed do have enough projections. Although we will work with the more general class of Rickart C*-algebras later on, since the former are much more familiar
it is instructive to first review the connection between commutative \vNa s and classical propositional logic. In the latter direction, let us recall the {\it Stone representation theorem} (see  \cite[\it passim]{johnstone82} or \cite[\S IX.10]{MLM}):
\begin{quote}
 Any Boolean lattice $\CL$  is  isomorphic to the lattice $\CB(\hat{\Sg}_{\CL})$ of clopen subsets of a
 {\it Stone space} $\hat{\Sg}_{\CL}$, i.e.,  a compact Hausdorff space that is {\it totally} disconnected,  in that the only connected subsets of $\hat{\Sg}_{\CL}$ are single points. (Equivalently, a Stone space is compact, $T_0$, and has a basis of clopen sets.)
 \end{quote}
Here
$\hat{\Sg}_{\CL}=\mathrm{pt}(\CL)$, called the {\it Stone spectrum} of $\CL$,  arises as the space  of `points' of $\CL$, which by definition are homomorphisms $\phv: \CL\raw \{0,1\}$ of  Boolean lattices (where $\{0,1\}\equiv\{\bot,\top\}$ as a lattice, i.e.\ $0\leqslant 1$ and $0\neq 1$),  topologized by declaring that the basic open sets in $\hat{\Sg}_{\CL}$
are those of the form $U_x=\{\phv \in \hat{\Sg}_{\CL}\mid \phv(x)=1\}$, for each $x\in\CL$. Such `points' $\phv\in\hat{\Sg}_{\CL}$ may be identified with maximal ideals\footnote{In this usage, an ideal $I$ in a lattice $L$ denotes a subset of $L$ such that $x,y\in I$ implies $x\vee y\in I$, and $y\leqslant x\in I$ implies $y\in I$. In a Boolean lattice, prime ideals and maximal ideals coincide, so that the Stone spectrum of a Boolean lattice is often described as the space of its prime ideals (which are those ideals that not contain 1 and where $x\wed y\in I$ implies either $x\in I$ or $y\in I$).} $I_{\phv}=\phv\inv(\{0\})\subset \CL$, topologized by saying that each $x\in \CL$ defines a basic open consisting of all maximal ideals not containing $x$. 
As in \er{inzicht}, one has a direct description of this topology as a frame (up to frame isomorphism), which turns out to be given by
\begin{equation}
\CO(\hat{\Sg}_{\CL})\cong  \Idl(\CL); \label{StoneJ}
\end{equation}
see Corollaries II.4.4 and II.3.3 and Proposition II.3.2 in \cite{johnstone82}. 

The following result describes the relationship between Boolean lattice and von Neumann algebras:\footnote{More generally, the proposition holds for Rickart C*-algebras, with the same proof.} \begin{proposition}\label{prop:vna}
Let $A$ be a von Neumann algebra. The following
conditions are equivalent:
\begin{enumerate}
\item $A$ is commutative;
\item The lattice $\CP(A)$ of projections in $A$ is Boolean.
 \end{enumerate}
 In that case, the Gelfand spectrum $\Sg_A$ of $A$ is homeomorphic to (and hence may be identified with) with the Stone spectrum $\hat{\Sg}_{\CP(A)}$
 of $\CP(A)$, and $\CP(A)$
 is isomorphic with the Boolean lattice $\CB(\Sg_A)$ of clopens in $\Sg_A$.  \end{proposition}
 \begin{proof}
 For the equivalence between 1 and 2 see \cite[Prop.\ 4.16]{Redei}. The homeomorphism 
 \begin{equation}
 \Sg_A\cong \hat{\Sg}_{\CP(A)} \label{homeo}
\end{equation}
 is clear from \er{inzicht} and \er{StoneJ}. The isomorphism of Boolean lattices
 \begin{eqnarray}
\CP(A) &\stackrel{\cong}{\raw}& \CB(\Sg_A); \label{isomor} \\
p&\mapsto & \CD(\hat{p}) \label{Piso}
\end{eqnarray}
then follows from  Stone's Theorem.\qed
 \end{proof}
  Here and in what follows,  for any $a\in C(\Sg_A)$ we write
\beq \label{defCD}
 \CD(a)=\{\sg\in\Sg\mid a(\sg)\neq 0\}. \eeq 
The homeomorphism \er{homeo} arises as follows:
\begin{itemize}
\item 
each  character $\phv:A\raw\C$, $\phv\in\Sg_A$,  restricts to a point $\hat{\phv}: \CP(A)\raw \{0,1\}$, $\hat{\phv}\in\hat{\Sg}_{\CP(A)}$;
\item conversely, each $\hat{\phv}\in\hat{\Sg}_{\CP(A)}$ extends to a character  $\phv\in\Sg_A$  by the spectral theorem. 
 \end{itemize}

 Proposition \ref{prop:vna} suggests that the projection lattices $\CP(A)$ of  general \vNa s may be seen as  noncommutative generalizations of classical propositional logic (in its semantic guise of Boolean algebras).  Despite the conceptual drawbacks we mentioned in Section \ref{ss1.2}, this gives a clear mathematical status to quantum logic in the style of Birkhoff and von Neumann. However, for various technical reasons the class of \vNa s is not optimal in this respect.  First, Proposition \ref{prop:vna} does not identify the class of Boolean lattices with the class of commutative \vNa s; in fact, if $A$ is a  commutative \vNa, then the lattice $\CP(A)$ is complete, so that $\Sg_A$ is not merely Stone but {\it Stonean}, i.e.\
 compact, Hausdorff and {\it extremely}  disconnected, in that the closure of every open set is open (and hence clopen).\footnote{The  Stone spectrum of a Boolean lattice $\CL$ is Stonean iff  $\CL$ is complete.}  But one does not obtain an identification of complete Boolean lattices (or, equivalently, Stonean spaces) with commutative \vNa s either, since the Gelfand spectrum
 of a commutative \vNa\ is not merely Stonean but has the stronger property
  of being {\it hyperstonean}, in admitting sufficiently many positive normal measures \cite[Def.\
1.14]{Tak1}. This is the situation: a commutative C*-algebra is a \vNa\ iff its Gelfand spectrum
(and hence the Stone spectrum of its projection lattice) is hyperstonean. Second, our use of
constructive mathematics in the main body of this paper leads to certain difficulties with the class
of \vNa s, mainly because they are defined on a given Hilbert space (as opposed to an abstract
\ca).\footnote{
Sakai's abstract characterization of \vNa s as C*-algebras that are the dual of some Banach space obviates this problem, but introduces others (notably the problem of internalizing the so-called ultraweak or $\sg$-weak topology on a \vNa), which we are unable to deal with constructively at the moment. A constructive theory of von Neumann algebras actually exists
\cite{Bridges/V:1999,Spitters:operator}, but this  theory relies on the use of
the strong operator topology, which has awkward continuity properties (e.g.,
the map $s\mapsto E_s$, where $E_s$ is the spectral projection associated to
$(-\infty,s)$, need not be strongly continuous).
 Furthermore, it uses the
 axiom of dependent choice, which  although available in our presheaf
topos defined below, is not valid in arbitrary toposes in which C*-algebras can be defined.}

To survey the landscape, we mention the basic classes of C*-algebras that are potentially relevant to logic in having sufficiently many projections, in order of increasing generality:\footnote{These definitions were originally motivated by the desire to find a purely algebraic
analogue of  the theory of \vNa s, rather than by quantum logic.}
\begin{definition}\label{defRCA}
A unital C*-algebra $A$ is said to be:
\begin{enumerate}
\item a \emph{\vNa} if it is the dual of some Banach space \cite{Sakai};
\item an \emph{$AW^*$-algebra}  if  for each nonempty subset $S\subset A$ there is a projection $p\in A$ so that
$R(S)=pA$ \cite{Kaplansky};
\item a \emph{Rickart \ca} if  for each  $x\in  A$ there is a projection $p\in A$ so that
$R(x)=pA$ \cite{Rickart46}; \label{R3}
\item a \emph{spectral \ca} if for each $a\in A$, $a\geq 0$, and each $\lm,\mu\in (0,\infty)$, $\lm<\mu$, there exists a projection $p\in A$ so that $ap\geq \lm p$ and $a(1-p)\leq \mu(1-p)$
\cite{SZ}.
\end{enumerate}
\end{definition}
 Here the {\it right-annihilator} $R(S)$ of $S\subset A$ is defined
as $R(S)=\{a\in A\mid xa=0\, \forall x\in S\}$ and $R(x)\equiv R(\{x\})$; in view of the presence
of an involution, equivalent definitions may be given in terms of the left-annihilator.
In all cases,  the projection $p$ is unique. It is known that if a C*-algebra $A$ has a faithful representation on a separable Hilbert space, then it is a Rickart C*-algebra iff it is an $AW^*$-algebra, but otherwise these classes are different.\footnote{It is generally believed that a \ca\
is Rickart iff it is  monotone $\sg$-complete. In that case, one may also define
 a  C*-algebra $A$ to be Rickart if each maximal Abelian $\mbox{}^*$-subalgebra of $A$ is monotone $\sg$-complete
 \cite{SaitoWright}.}  Let us note that the equivalence between the original definition of a \vNa\ and the one given here is quite a deep result in the theory of operator algebras.

 We now have the following results, of which the first has already been mentioned. Recall that $\CB(\Sg)$ is the Boolean lattice of clopens of a Stone space $\Sg$; as in the case of \vNa s, if $\Sg_A$ is the Gelfand spectrum of a commutative C*-algebra $A$, then
 $\CB(\Sg_A)$ is isomorphic to the lattice $\CP(A)$ of projections in $A$.
 \begin{theorem} Let $A$ be a commutative C*-algebra with Gelfand spectrum $\Sg_A$. Then $A$ is:
 \begin{enumerate}
\item a \vNa\ iff $\Sg_A$ is hyperstonean  \cite[\S III.1]{Tak1};
\item an $AW^*$-algebra iff $\Sg_A$ is Stonean (equivalently, Stone  with the additional property that $\CB(\Sg_A)$ is complete) \cite[Thm.\ 1.7.1]{Berberian};
\item a Rickart C*-algebra iff $\Sg_A$ is Stone with the additional property that  $\CB(\Sg_A)$ is $\sg$-complete \cite[Thm.\ 1.8.1]{Berberian};
\item a spectral C*-algebra iff $\Sg_A$ is Stone \cite[\S 9.7]{SZ}.
\end{enumerate}
Restricting Gelfand duality to each of the above cases results in a
categorical duality (\eg,  for case 3 above, between commutative Rickart C*-algebras and
Stone spaces $X$ for which $\CB(X)$ is $\sigma$-complete).
\end{theorem}
The  completeness of $\CB(\Sg)$ is equivalent to the property that the closure of the union of any family  of clopens in $\Sg$ is clopen;  similarly, $\CB(\Sg)$  is $\sg$-complete iff the closure of the union of a countable family  of clopens in $\Sg$ is clopen.

It appears that in the commutative case spectral C*-algebras form the most general class to work with from the point of view of classical logic, but unfortunately, the projections in a noncommutative spectral C*-algebra may not form a lattice. A major advantage of Rickart C*-algebras is that they do \cite[Prop.\ 1.3.7 and Lemma 1.8.3]{Berberian}:
\begin{proposition}\label{eigRCA}
The set of projections $\CP(A)$ in a Rickart C*-algebra  $A$ form a $\sg$-complete lattice under the ordering $p\leq q$ iff $pA\subseteq qA$.
\end{proposition}
The ensuing lattice operations are given by
\begin{eqnarray}
p\wed q&=& q+  \mathrm{RP}[(p(1-q)];
 \label{infR}\\
p\vee q&=& p- \mathrm{LP}[(p(1-q)],
 \label{supR}
\end{eqnarray}
where for any $x\in A$ the projections $\mathrm{RP}[x]$ and $\mathrm{LP}[x]$ are defined by
$R(x)=(1-\mathrm{RP}[x])A$ and $L(x)=A(1-\mathrm{LP}[x])$, respectively (i.e., $\mathrm{RP}[x]=1-p$
where $R(x)=pA$, etc.).
We also have properties that guarantee the availability of spectral theory  (the strong limits in the usual constructions are just replaced by limits of monotone positive sequences):
\begin{proposition}
\begin{enumerate}
\item A commutative Rickart C*-algebra  is the (norm-)closed linear span of its projections \cite[Prop.\ 1.8.1.(3)]{Berberian};
\item  A commutative Rickart C*-algebra  $C$ is monotone $\sg$-complete, in that each increasing bounded sequence  of self-adjoint elements of $C$ has a supremum in $C$ \cite[Prop.\ 9.2.6.1]{SZ}.\footnote{Quoted in \cite[p.\ 4728]{DAntonioZsido}.
Similarly, a commutative $AW^*$-algebra is monotone complete.
It is an open question whether any  Rickart C*-algebra  $C$ is monotone $\sg$-complete.}
\end{enumerate}
\end{proposition}

In our search for  a suitable class of operator algebras to lie at the basis of intuitionistic quantum logic, and in particular to generalize the Heyting algebra (or frame) \er{bohr} to all elements $A$ of this class, we also require certain constructions to work internally in a topos; in particular, the ``Bohrification'' $\uA$ of $A$ (defined in the next section) should internally lie in the same class as $A$ itself. This will indeed be the case for  Rickart C*-algebras; see Theorem \ref{intRick} below. Summing up,
we generalize the usual algebraic approach to quantum logic \cite{Redei} in proposing that instead
of \vNa s, we prefer to work with
Rickart C*-algebras. All one loses in this generalization is the completeness of the projection lattice $\CP(A)$ of $A$, but since one does have the slightly weaker property of $\sg$-completeness (which, if $A$ has a faithful representation on a separable \Hs, actually implies the completeness of  $\CP(A)$), this is not a source of tremendous worry.
\section{Internal  Rickart C*-algebras}\label{RA1}
In this section we assume familiarity with basic category and topos theory;
the Appendix to \cite{caspersheunenlandsmanspitters:nlevelsystem} is tailor-made for this purpose, and also the first few chapters of \cite{ML} and \cite{MLM} contain all necessary background. See also \cite{Bell,Gold} for introductions that emphasise
 the connection between topos theory and intuitionistic logic. In some technical arguments we will also use the so-called internal language of a topos and its Kripke--Joyal semantics, for which \cite[Ch.\ VI]{MLM} is our basic reference.
 Briefly, a topos may be seen as a generalization of the category \Sets\ (whose objects are sets and whose arrows are functions, subject to the usual ZFC axiom system) in which
most set-theoretic reasoning can be carried out, with the restriction
that all proofs need to be constructive in the limited sense that one
cannot make use of the law of the excluded middle or the Axiom of
Choice. In what follows, we will use the term `constructive' in this
way.\footnote{The reader be warned that topos theory makes extensive
use of the power set construction, which is avoided in
so-called predicative constructive mathematics.}

Let $A$ be a Rickart \ca, with associated poset $\CA$ of all unital commutative Rickart C*-subalgebras of $A$,
partially ordered by set-theoretic inclusion. The poset $\CA$ defines a category, called $\CA$ as well, in which
$C$ and $D$ are connected by a unique arrow $C\raw D$ iff $C\subseteq D$, and are not connected by any arrow otherwise.
In this paper, the only relevant  topos besides \Sets\ is the category
\beq
\TA=\Sets^{\CA}\label{defTA}
\eeq
of (covariant) functors from $\CA$, seen as a category, to \Sets.  We will $\underline{\mathrm{underline}}$  objects in $\TA$. As a case in point,   the tautological functor
\beq \ulA:C\mapsto C, \label{intA}\eeq
maps a point $C\in\CA$  to the corresponding commutative C*-algebra $C\subset A$ (seen as a set);
for $C\subseteq D$ the map $\uA(C\leq D):\uA(C)\raw \uA(D)$ is just the inclusion $C\hookrightarrow D$.
We call $\ulA$ the {\it Bohrification} of $A$.
\begin{theorem}\label{intRick}
Let $A$ be a Rickart \ca. Then $\uA$ is a commutative Rickart C*-algebra in $\TA$.
\end{theorem}
\begin{proof}
Since $A$ is, in particular, a \ca, it follows from \cite[Thm.\
5]{HLS} that $\uA$ is a commutative C*-algebra in $\TA$.
To prove that it is internally Rickart, we spell out Definition \ref{defRCA}.\ref{R3} in logical notation,
 with $x\in A$ as a free variable:
\beq \exists_{p\in A}\, xp=0\wed \forall_{y\in A}\, xy=0\Raw  y=py. \label{logdef}\eeq
Here we have changed the condition inherent in  Definition \ref{defRCA}.\ref{R3}  that $xy=0$ implies that there exists $a\in A$ such that $y=pa$, to the equivalent condition that $xy=0$ implies $y=py$; see \cite[Prop.\ 1.3.3]{Berberian}. This is not necessary, but simplifies the argument somewhat.

We regard  \er{logdef} as a formula $\phi$ in the internal language of $\TA$ with a single free variable $x$ of type
$\uA$. By Kripke--Joyal semantics, $\phi$ is true if $C\Vdash\phi(\til{x})$ for all $C\in\CA$ and all $\til{x}\in \uA(C)=C$ \cite[\S VI.7]{MLM}. By the rules for this semantics, $C\Vdash\ph(\til{x})$ is true iff there exists a projection $\til{p}\in C$ such that
for all $D\supseteq C$, all $\til{y}\in D$, and all $E\supseteq D$ one has: if $\til{x}\til{y}=0$, then $\til{y}=\til{p}\til{y}$.
In the latter part, the elements $\til{x}\in C$, $\til{p}\in C$, and $\til{y}\in D$ are all regarded as elements of $E$, but
clearly the if \ldots then statement holds at all $E\supseteq D$ iff it holds at $D$. The truth of $C\Vdash\phi(\til{x})$, and hence of Theorem \ref{intRick} now follows from the following lemma.
\begin{lemma}\label{lemR}
Let $C$ and $D$ be commutative Rickart C*-algebras with $C\subseteq D$, and take $x\in C$. If one regards $x$ as an element of $D$, then the projection $p$ for which $R(x)=pD$  lies in $C$. In other words: if $x\in C\subseteq D$, then  the projection $\mathrm{RP}[x]$ as computed in $D$ actually lies in $C$.
\end{lemma}
\begin{proof}
We have $C\cong C(\Sg_C)$ and $D\cong C(\Sg_D)$ through the Gelfand transform. As we have seen, $\Sg_C$ is a Stone space, whose topology has a basis $\CB(\Sg_C)$ consisting of all clopen sets in $\Sg_C$. This basis is isomorphic as a Boolean lattice to the projection lattice $\CP(C)$ of $C$, with isomorphism \er{Piso} (for $A=C$), and
analogously for $D$.

One has a canonical map $r_{DC}:\Sg_D\raw\Sg_C$ given by restriction, i.e.\ $(r_{DC}\phv)(a)=\phv(a)$ for $a\in C$, or $r_{DC}\phv=\phv_{| C}$. Being continuous, this map induces the inverse image map
\beq r_{DC}\inv: \CO(\Sg_C)\raw\CO(\Sg_D) \label{rDC}
\eeq
 as well as the pullback
\beq
r_{DC}^*: C(\Sg_C)\raw C(\Sg_D).
\eeq Restricted to basic opens and projections, respectively, these maps are related to each other and to the inclusion $\iota_{CD}:C\hookrightarrow D$ by
\begin{eqnarray}
  r_{DC}^*(\hat{p}) &=& \widehat{\iota_{CD}(p)}; \label{rel1} \\
 r_{DC}\inv(\CD(\hat{p}))&=& \CD( r_{DC}^*(\hat{p})).\label{rel2}
\end{eqnarray}
By \cite[Prop.\ 1.8.1.(4)]{Berberian}, the projection $p\in D$ in the statement of the Lemma
has Gelfand transform  $\hat{p}=1-\ch_{\CD(r_{DC}^*x)^-}$, where for any $U\subset\Sg$,
$U^-$ is the closure of $U$.
 But by  \er{rel2} one then has $\hat{p}=r_{DC}^*(\hat{q})$ with
  $\hat{q}=(1-\ch_{\CD(x)^-})$, and \er{rel1} yields $p=\iota_{CD}(q)$.
Hence $p\in C$.
This concludes the proof of Lemma \ref{lemR} as well as of Theorem \ref{intRick}.\qed
\end{proof}\end{proof}

%  Theorem \ref{intRick}, or perhaps the desire to develop as much of  C*-algebra theory in a topos
% as % possible, motivates the constructive study of Rickart C*-algebras, particularly of
% commutative ones. % Any result obtained in this way may then be used internally, i.e.\ in an
% arbitrary topos.

Wa now initiate  a constructive theory
of Rickart C*-algebras. Our constructive approach is crucial for what follows, for any constructive result may be used internally, i.e.\ in an arbitrary topos. In addition, it leads to an
 alternative proof of Theorem~\ref{intRick}, which may be rederived from the Proposition~\ref{prop:unique} below.\footnote{
 Proposition~\ref{prop:unique}
 shows that Rickart C*-algebras are C*-algebras
equipped with an extra (partial) operation $a\mapsto [a>0]$. A proof of Theorem~\ref{intRick} may then be
obtained by a simple extension of~\cite[Thm.~5]{HLS} by observing that the definition of f-algebras
with such an operation is Cartesian and hence geometric.}
 \begin{proposition}\label{prop:unique}
Let $A$ be a commutative C*-algebra. The following are equivalent:
\begin{enumerate}
 \item for each $a\in A$ there exists a (unique) projection $p$ such that i) $ap=0$
and ii)
if $ab=0$, then there exists $c$ such that $b=cp$.
 \item for each $a$ there exists a (unique) projection $p$ such that $ap=0$ and
if $ab=0$, then $b=bp$.
 \item for each self-adjoint $a$ there exists a (unique) projection, denoted
$\groter{a}{0}$,
such that $\groter{a}{0}a=a^+$ and $\groter{a}{0}\wedge
\groter{-a}{0}=0$.\\
\end{enumerate}
\end{proposition}
Let us note that since $A$ is commutative, the infimum $\wed$ in 3 is the same as the product.
\begin{proof}
The equivalence of 1 and 2 is in~\cite[Prop 1.3.3]{Berberian}.
We denote the projection $p$ in 2 by $[a=0]$. By the decomposition of arbitrary
elements of $A$ in four positives, it suffices to require the
existence of $[a=0]$ only for positive elements $a$; for general $a\in A$
we obtain the required projection by multiplication of the four projections for its positive components.
\begin{itemize}
 \item[2$\to$3] For a self-adjoint $a$ we define
$\groter{a}{0}:=1-[a^+=0]$. Then
\[\groter{a}{0}a=(1-[a^+=0])(a^+-a^-)=a^+.\]
By definition, $\groter{a}{0}= \groter{a^+}{0}$. By 2 and $a^-a^+=0$,
$a^-\groter{a}{0}=0$. Again by 2, but applied to $a^-$,
$\groter{a^-}{0}\groter{a}{0}=0$. Since $(-a)^+=a^-$, $\groter{a}{0}\wedge
\groter{-a}{0}=0$.
\item[3$\to$2] For positive $a$ we define
$[a=0]:=1-\groter{a}{0}$. Then
$a[a=0]=a(1-\groter{a}{0})=0$. We may assume that $a,b\geq 0$ and
$ab=0$. Then  $b\groter{a}{0}\leq 0$ (see part 1 of Lemma
\ref{lem:multiplication} below),
and since $b\groter{a}{0}$ is the product of commuting positive operators, this implies
$b\groter{a}{0}=0$. \qed
\end{itemize}
\end{proof}

Thus $A$ is a commutative Rickart C*-algebra if any (and hence all) of the three conditions in this proposition is satisfied.
Our earlier proof of Theorem~\ref{intRick} can now be reformulated in a simple way by applying the above proposition to
$\uA$:  since the existence of the projection
$[a>0]$ in part 3 of Proposition~\ref{prop:unique} is interpreted locally, $\uA$ satisfies the condition in 3 if each
 $C\in\CA$ does. Hence $\uA$ is Rickart.

Similarly, the $\sg$-completeness of the projection lattice of a commutative Rickart C*-algebra (cf.\ Proposition~\ref{eigRCA})
is immediate from the following analogue of \cite[Lem.~1.8.2]{Berberian}:
\begin{lemma}\label{lem:Berb}
 A sequence $p_n$ of mutual orthogonal projections has a supremum.
\end{lemma}
\begin{proof}
 The sum $a:=\sum 2^{-n}p_n$ converges in the C*-algebra. The supremum of the
sequence is the projection $[a>0]$.\qed
\end{proof}

Finally, Sait\^o and
Wright~\cite{SaitoWright} define a C*-algebra to be Rickart  if each maximal Abelian
*-subalgebra of $A$ is Rickart (or, equivalently, monotone $\sg$-complete).
Equivalently, one may require that every Abelian *-subalgebra is contained in an
Abelian Rickart C*-algebra.
This definition captures essential parts of the theory of von
Neumann algebras  and, being formulated entirely in terms of commutative subalgebras, is very much in the spirit of our ``Bohrification'' program. Unfortunately, although
every Rickart C*-algebra in the sense of Definition \ref{defRCA} is a Rickart \ca\
in the sense of Sait\^o and Wright,  the converse has not been shown to date.\footnote{Private communications from  Sait\^o and Wright.} In any case, upon the
definition of  Sait\^o and Wright, Rickart C*-algebras admit a nice internal characterization, provided
we use classical meta-logic and use the original definition of the poset $C(A)$ from~\cite{HLS}, according to which
$\CA$ is the collection of {\it all} commutative unital C*-subalgebras of $A$.
\begin{proposition} \label{SW}
Let $A$ be a C*-algebra in $Sets$. Then $A$ is a Rickart C*-algebra in the sense of Sait\^o and Wright iff $\uA$
satisfies: for all self-adjoint $a$, not not there exists a projection $p$ such
that $p=\groter{a}{0}$.
\end{proposition}\begin{proof}
By Lemma~\ref{eventually} below, the right hand side means that for all $D\in
\CA$ and $\til{a}\in D\sa$ there exists $E\supset D$ and a projection $\til{p} \in E$ such
that $E\Vdash (p=\groter{a}{0})(\til{p},\til{a})$, i.e.\ $\til{p}=\groter{\til{a}}{0}$ in $E$. This is precisely our earlier definition of a Rickart \ca.
\qed\end{proof}
The use of Proposition \ref{SW} derives from the fact that in both classical and intuitionistic logic, the propositions $A\to\neg B$ and $\neg\neg A\to \neg B$ are equivalent. Hence
negative statements for Rickart algebras may be proved by assuming that the
projection $\groter{a}{0}$ actually exists.
\section{Gelfand theory for commutative Rickart C*-algebras}\label{RA2}
The Gelfand theory for commutative C*-algebras $A$
 that in the classical case leads to the isomorphism $A\cong C(\Sg_A)$ for some compact Hausdorff space $\Sg_A$, generalizes
 to the constructive or topos-theoretic setting in producing a frame (see Section \ref{sec4})  $\CO(\Sg_A)$, rather than the space $\Sg$ itself, with the property that $A$ is isomorphic as a commutative C*-algebra with the object of all frame maps from $\CO(\C)$
 (i.e.\ the frame of Dedekind complex numbers, interpreted in the ambient topos) to  $\CO(\Sg_A)$. In the classical case,
 since $\Sg_A$ is Hausdorff and hence sober,
 each frame map $\phv^*:\CO(\C)\raw\CO(\Sg_A)$ arises as the inverse image $\phv^*=\phv\inv$ of some continuous map
 $\phv:\Sg_A\raw\C$, so that one recovers the usual Gelfand isomorphism, but in general this isomorphism involves the frame $\CO(\Sg_A)$ in the said way; an underlying space $\Sg_A$ may not even exist (indeed, due to the Kochen--Specker Theorem this is precisely the case in our application to quantum theory).\footnote{The notation $\CO(\Sg_A)$ for a frame whose underlying point set $\Sg_A$ may not exist may appear odd, but is generally used in order to stress that one may reason with $\CO(\Sg_A)$ as if it were the topology of some space.}

 The abstract theory of internal C*-algebras and Gelfand duality in a topos is due to Banaschewski and Mulvey \cite{BM}.
 In order to explicitly compute the frame $\CO(\Sg_A)$ for given $A$,
 we use the constructive formulation of Gelfand duality  due to
 Coquand and Spitters~\cite{coquandspitters05,CoquandSpitters:cstar}, building on fundamental
insights into
 Stone duality by Coquand \cite{Coq}; see also \cite{HLS}.
 First, define a relation $\preccurlyeq$ on the self-adjoint part $A\sa=\{a\in A\mid a^*=a\}$ of $A$ by putting
$a\preccurlyeq b$ iff there exists an
$n\in \N$ such that $a\leq n b^+$. This yields an associated equivalence relation
$a\approx b$, defined by $a \preccurlyeq b$ and
$b \preccurlyeq a$.
The lattice $L_A$ is defined
as
\beq L_A= A^+/\approx ,\eeq
where $A^+=\{a\in A\mid a\geq 0\}$ is the positive cone of $A$.

The key results are that $L_A$ is a  so-called normal distributive lattice and that $\CO(\Sg_A)$ arises as the frame
$\mathrm{RIdl}(L_A)$ of regular ideals
in $L_A$. We shall not define these notions here (see  \cite{Coq,coquandspitters05,HLS}), since in the case at hand the situation simplifies according to Theorem \ref{RCAS} below, but we will need the following information.
We denote the equivalence class of $a\in A\sa$ in $L_A$ by $D(a)$; we have $D(a)=D(a^+)$, so that we may restrict $a$ to lie in $A^+$, i.e.\ $a\geq 0$. Furthermore, we denote  the map $L_A\to \mathrm{RIdl}(L_A)$ that assigns the regular closure of the principal down set
$\downarrow\! D(a)$ to $D(a)\in L_A$ (see \cite[Thm.\ 27]{coquand:entail} or \cite[eq.\ (80)]{HLS})
by $D(a)\mapsto\CD(a)$;
upon the identification $\CO(\Sg_A)\cong  \mathrm{RIdl}(L_A)$, this map simply injects $D(a)$ into $\CO(\Sg_A)$ as a basis open, and in the classical case this notation is consistent with \er{defCD}. On then has the following relations:
\begin{eqnarray}
\CD(1)  & =& \top; \label{eq:spectrum1} \\
  \CD(a)\wedge \CD({-a}) & =& \bot; \label{eq:spectrum2} \\
                \CD({-b^2}) & =& \bot;  \label{eq:spectrum3} \\
             \CD({a+b}) & \leqslant&  \CD(a)\vee
                                  \CD(b), \label{eq:spectrum4} \\
              \CD({ab}) & =&  (\CD(a) \wedge \CD(b)) \vee
                                    (\CD({-a}) \wedge
                                    \CD({-b})), \label{eq:spectrum5}\\
  \CD({a}) &= & \bigvee_{s>0}\CD({a-s}).
    \label{eq:continuity}
\end{eqnarray}
In fact, the first five relations already hold for the $D(\cdot)$ and
may be used to define $L_A$, whereas the complete set may be used as a
definition of $\CO(\Sg_A)$.

We now work towards the explicit formula for the external description
of the Gelfand spectrum of the Bohrification of a Rickart C*-algebra
in Theorem~\ref{RCAS} below.

\begin{lemma}
\label{lem:multiplication}
  Let $A$ be a commutative Rickart C*-algebra, and $a,b \in A$
  self-adjoint. If $a \leq ab$, then $a \preccurlyeq b$, \ie\ $D(a)
  \leq D(b)$.
\end{lemma}
\begin{proof}
  If $a \leq ab$ then certainly $a \preccurlyeq ab$. Hence $D(a) \leq
  D(ab) = D(a) \wedge D(b)$. In other words, $D(a) \leq D(b)$,
  whence $a \preccurlyeq b$.
  \qed
\end{proof}

\begin{definition}
 \cite{Gold,johnstone82}
  A \emph{pseudocomplement} on a distributive lattice $L$ is an
  antitone (i.e.\ anti-monotone) function $\neg \colon L \to L$ satisfying $x \wedge y=0$
  iff $x \leq \neg y$.\footnote{The construction of the Boolean
  algebra of projections as the pseudocomplements in the lattice $L$
  is reminiscent of the construction of the Boolean algebra of
  pseudocomplements which can be carried out in a Heyting algebra;
  e.g.~\cite[I.1.13]{johnstone82}. However, as $L$ need not be a
  Heyting algebra, our construction is not an instance of this general
  method.}
\end{definition}

\begin{proposition}
  For a commutative Rickart C*-algebra $A$, the lattice $L_A$ has a
  pseudocomplement, determined by $\neg D(a) = D([a=0])$ for $a \in
  A^+$.
\end{proposition}
\begin{proof}
  Without loss of generality, let $b \leq 1$. Then
  \begin{align*}
           D(a) \wedge D(b) = 0
    & \iff D(ab) = D(0) \\
    & \iff ab = 0 \\
    & \iff b[a=0]=b \eqcomment{($\Rightarrow$ by Proposition~\ref{prop:unique})} \\
    & \iff b \preccurlyeq [a=0] \eqcomment{($\Leftarrow$ since $b \leq 1$,
      $\Rightarrow$ by Lemma~\ref{lem:multiplication})} \\
    & \iff D(b) \leq D([a=0]) = \neg D(a).
  \end{align*}
  To see that $\neg$ is antitone, suppose that $D(a) \leq D(b)$. Then
  $a \preccurlyeq b$, so $a \leq nb$ for some $n \in \field{N}$. Hence
  $[b=0]a \leq [b=0]bn=0$, so that $\neg D(b) \wedge D(a) = D([b=0]a)
  = 0$, and therefore $\neg D(b) \leq \neg D(a)$.
  \qed
\end{proof}

\begin{lemma}
\label{lem:regularityruleRickart}
  If $A$ is a commutative Rickart C*-algebra, then the lattice $L_A$
  satisfies $D(a) = \bigvee_{r \in \field{Q}^+}
  D([a-r>0])$ for all $a \in A^+$.
\end{lemma}
%Compare the above to~\eqref{eq:continuity}.
\begin{proof}
  Since $[a>0]a = a^+ \geq a$, Lemma~\ref{lem:multiplication} gives
  $a \preccurlyeq [a>0]$ and therefore $D(a) \leq D([a>0])$.
  Also, for $r \in \field{Q}^+$ and $a \in A^+$, one has
  $1 \leq \frac{2}{r}((r-a) \vee
  a)$, whence
  \[
    [a-r>0] \leq \frac{2}{r}((r-a)\vee a)[a-r>0] = \frac{2}{r}(a[a-r>0]).
  \]
  Lemma~\ref{lem:multiplication} then yields $D([a-r>0]) \leq
  D(\frac{2}{r}a) = D(a)$.
  In total, we have $D([a-r>0]) \leq D(a) \leq D([a>0])$ for all
  $r \in \field{Q}^+$, from which the statement follows.
  \qed
\end{proof}

\begin{theorem}
\label{RCAS}
  The Gelfand spectrum $\CO(\Sg_A)$ of a commutative Rickart
  C*-algebra $A$ is isomorphic to the frame
  $\mathrm{Idl}(\CP(A))$ of ideals of $\CP(A)$.
\end{theorem}
\begin{proof}
  Form the sublattice $\CP_A = \{ D(a) \in L_A \mid a \in A^+, \neg\neg
  D(a) = D(a) \}$ of `clopen elements' of $L_A$, which is Boolean by
  construction. Since $\neg D(p) = D(1-p)$ for $p \in \CP(A)$, we
  have $\neg\neg D(p) = D(p)$. Conversely, $\neg\neg D(a) =
  D([a>0])$, so that each element of $\CP_A$ is of the form $D(a) = D(p)$
  for some $p \in \CP(A)$. So $\CP_A = \{ D(p) \mid p \in \CP(A)
  \} \cong \CP(A)$, since each projection $p \in \CP(A)$ may be
  selected as the unique representative of its equivalence class
  $D(p)$ in $L_A$. By Lemma~\ref{lem:regularityruleRickart},
  we may use $\CP(A)$ instead of $L_A$ as the generating lattice for
  $\CO(\Sg_A)$. So $\CO(\Sg_A)$ is the collection of regular
  ideals of $\CP(A)$ by \cite[Theorem~26]{HLS}. But since
  $\CP(A) \cong \CP_A$ is Boolean, all its ideals are regular, as
  $D(p) \ll D(p)$ for each $p \in
  \CP(A)$~\cite{johnstone82}. This establishes the
  statement.
    \qed
\end{proof}

Internalized to the topos $\TA$, Theorem \ref{RCAS} enables us to compute the spectrum $\CO(\uSg_{\uA})$ of the Bohrification $\uA$ of $A$.
As a functor $\CO(\uS_{\uA}):\CA\raw\Sets$, this spectrum is completely determined by its component at $\C\cdot 1$, which is the frame in \Sets\ that provides the so-called external description of $\CO(\uSg_{\uA})$ \cite{JT}
(see also \cite[Thm.\ 29]{HLS}).  We write
\beq
\CO(\Sg_A)\equiv \CO(\uS_{\uA})(\C\cdot 1),\eeq
and draw attention to the notation \er{rDC}.
\begin{theorem}\label{maintheorem}
The frame $\CO(\Sg_A)$ is given by
\beq \CO(\Sg_A)=\{S: \CA\raw\Sets\mid S(C)\in\CO(\Sg_C), r_{DC}\inv(S(C))\subseteq S(D) \mbox { if } C\subseteq D\}, \label{bohr3}
\eeq
and has a basis given by
\beq
\CB(\Sg_A)=  \{\tilde{S}:\mathcal{C}(A) \raw \CP(A)\mid {\tilde{S}}(C)\in \CP(C),\,
{\tilde{S}}(C)\leq {\tilde{S}}(D)\mbox { if } C\subseteq D\}, \label{bohr2}
\eeq
in the sense that under the (injective) map $f:\CB(\Sg_A)\raw  \CO(\Sg_A)$ given by
\beq f(\til{S})(C)=\CD(\widehat{\til{S}(C)}), \eeq
each $S\in  \CO(\Sg_A)$ may be expressed as $S=\bigvee\{f(\tilde{S})\mid \tilde{S}\in \CB(\Sg_A), f(\tilde{S})\leq S\}$.
 \end{theorem}
\begin{proof}
We interpret  Theorem \ref{RCAS} in the topos $\TA$, where $\uA$ plays the role of the general commutative C*-algebra $A$ in the above analysis (not to be confused with the noncommutative C*-algebra $A$ in \Sets\ whose Bohrification is $\uA$).
The internal version of the lattice $L_A$ is the functor $\underline{L}_{\uA}$, which according to \cite[Thm.\ 20]{HLS} is simply given by $\underline{L}_{\uA}(C)=C$. Consequently, the subobject $\underline{\CP}_{\uA}$ is given by
$\underline{\CP}_{\uA}(C)=\CP(C)$ (as the algebraic conditions $p^2=p^*=p$ defining a projection are interpreted locally).

Combining Theorem \ref{RCAS} with Theorem 29 in \cite{HLS}, we find that
\beq
\CO(\Sg_A)\cong \mathrm{Idl}(\underline{\CP}_{\uA}),\label{combi}
\eeq
where the right-hand side by definition is the subset of $\mathrm{Sub}(\underline{\CP}_{\uA})$ that consists of
subfunctors $\underline{U}$ of $\underline{\CP}_{\uA}$
for which
 $\underline{U}(C)\in \mathrm{Idl}(\CP(C))$ for all $C\in\CA$. Now, internalizing Theorem \ref{RCAS} to \Sets\ and applying it to $A=C$, we obtain the
frame isomorphism $\mathrm{Idl}(\CP(C))\cong\CO(\Sg_C)$; the identification is given by mapping $I\in \mathrm{Idl}(\CP(C))$ to $\bigcup\, \{\CD(\hat{p})\mid p\in I\} \in \CO(\Sg_C)$.
 The requirement that  $\underline{U}$ be a subfunctor of $\underline{\CP}_{\uA}$ then immediately yields \er{bohr3}.
 Part 2
is obvious from the fact that the order in $\CO(\Sg_A)$ and in $\CB(\Sg_A)$ is defined pointwise.
\qed
\end{proof}

Now let $A=M_n(\C)$. By the Kochen--Specker theorem in the version given in \cite{HLS} and \cite{caspersheunenlandsmanspitters:nlevelsystem}, the frame (more precisely, the locale)  $\CO(\uS_{\uA})$
does not have any point. In particular, it cannot have $n$ points. Classically, of course, one has
$\Sg_{\C^n}=\mathbf{n}\equiv \{1,2,\ldots, n\}$ and hence
\beq
\CO(\Sg_{\C^n})\cong\CP(\C^n)\cong  \Pow(\mathbf{n}) \label{hence}\eeq
 (i.e.\ the power set of
$\mathbf{n}$).\footnote{We use the notation $\Pow(X)$ for the power set of $X$ to distinguish it -- in a constructive setting --
from $2^X$, which is used to denote the set of \emph{decidable} subsets of $X$, i.e.\ subsets $Y$
such that for all $x$ in $X$, $x\in Y$ or $x\not\in Y$. In the presence of classical logic all
subsets are decidable, so that $2^X\equiv \Pow(X)$.} The points
of $\Sg_{\C^n}$ are in
bijective correspondence with the  completely prime filters of $\Pow(\mathbf{n})$, and hence, once again,  with the elements of $\mathbf{n}$.
 Remarkably, we can prove that it is not not the case that internally $\CO(\uS_{\uA})$ has precisely the same structure.
\begin{proposition}
Let  $A=M_n(\C)$. Then
it is impossible that the Gelfand spectrum $\uS_{\uA}$ does not have
$n$ points. More precisely, noting that  in $\TA$ the set $\mathbf{n}$ is internalized as the constant
functor  $\underline{\mathbf{n}}: C\mapsto \mathbf{n}$, we internally have
\begin{eqnarray}
\neg\neg\, (
\underline{\CP}_{\uA} &\cong& \underline{\Om}^{\mathbf{\underline{\mathbf{n}}}});\label{isoo1} \\
\neg\neg \, (\CO(\uS_{\uA}) & \cong &
\underline{\Omega}^{\mathbf{\underline{\mathbf{n}}}}).\label{isoo2}
\end{eqnarray}
\end{proposition}
\begin{proof}
The proof relies on the following lemma.
\begin{lemma}\label{eventually}
Let $\phi$ be a formula in the internal language of $\TA$ (for simplicity without free variables). Then
$C\Vdash \neg \neg \phi$ iff $\phi$ holds eventually, in that for all $D\supseteq C$ there
exists $E\supseteq D$ such that $E\Vdash \phi$. In particular, $\phi$ is true if $E\Vdash \phi$ for any maximal
commutative C*-subalgebra $E$ of $A$.
\end{lemma}
\begin{proof}
By Kripke--Joyal semantics, we have
$C\Vdash \neg \neg \phi$ iff for all
 $D\supseteq C$, not $D\Vdash \neg \phi$, which is the case  iff
 for all  $D\supseteq C$, not for all $E\supseteq D$ not
$D\Vdash \phi$. If we now use classical meta-logic, we have $\neg \forall_x\, \neg \phi(x)$ iff
$\exists_x\, \phi(x)$. Then
the last condition holds iff
for all  $D\supseteq C$ there exists $E\supseteq D$ such that $E\Vdash \phi$.
\qed\end{proof}

The  power set
$\Pow(\mathbf{n})$ internalizes as the functor
\[\uOm^{\underline{\mathbf{n}}}: C \mapsto \mathrm{Sub}(\underline{\mathbf{n}}_{|\uparrow C}),\]
where the right-hand side is the set of all subfunctors of the functor
$\underline{\mathbf{n}}$ truncated to $\uparrow\! C\subset\CA$. In particular, if $E$ is a maximal
commutative C*-subalgebra of $M_n(\C)$, using \er{hence} and $\dim(E)=n$  we have
\beq
 \uOm^{\underline{\mathbf{n}}}(E)= \mathrm{Sub}(\underline{\mathbf{n}}_{|E})\cong \CP(E)\cong
\Pow(\mathbf{n})\label{this}\eeq
as (Boolean) lattices in \Sets. We now show that we may rewrite \er{this}  as $E\Vdash\underline{\CP}_{\uA}\cong \underline{\Om}^{\mathbf{\underline{n}}}$.
Namely,  $\mathcal{P}(E)\cong \Pow(\mathbf{n})$ iff there are   $f: \Pow(\mathbf{n})\raw \CP(E)$
and $g: \CP(E)\raw \Pow(\mathbf{n})$
such that
 $f(g(p))=p$ for all $p\in \mathcal{P}(E)$ and
$g(f(Y))=Y$  for all $Y\in \Pow(\mathbf{n})$.
 Now
$E \Vdash \forall p \in \mathcal{P}. f(g(p))=p$ iff
for all $F\supseteq E$ and $p$ in $\mathcal{P}(E)$,
$F \Vdash f(g(p))=p$. Since $E$ is maximal this is just:
for all $p$ in $\mathcal{P}(E)$,
$E \Vdash f(g(p))=p$, which is true.
Similarly, $E\Vdash g\circ f=id$.
Lemma~\ref{eventually} then gives
$C\Vdash \neg\neg\, (
\underline{\CP}_{\uA} \cong \underline{\Om}^{\mathbf{\underline{\mathbf{n}}}})$  for each $C\in\CA$, and hence \er{isoo1}.

We now show that this implies \er{isoo2}. Indeed,
to prove $\neg \neg A\to \neg \neg B$ it suffices to show that $A\to B$, so that {\it for the purpose of proving
 \er{isoo2}} we may assume $\underline{\CP}_{\uA} \cong \underline{\Om}^{\mathbf{n}}$.
By Theorem \ref{RCAS}, one then has  $\CO(\uS_{\uA})\cong
\mathrm{Idl}(\underline{\CP}_{\uA})\cong
\mathrm{Idl}(\underline{\Om}^{\mathbf{\underline{\mathbf{n}}}})\cong \underline{\Omega}^{\mathbf{\underline {n}}}$ (where the last isomorphism is most easily proved internally).
\qed\end{proof}
\section{Partial Boolean algebras and Bruns--Lakser completions}\label{secor}
This section compares the construction of our (complete) Heyting algebra
$\CO(\Sg_A)$ of Theorem~\ref{maintheorem} to some more
traditional descriptions of the logical structure
of quantum-mechanical systems, notably as far as distributivity and implication are involved.
Furthermore, we compare our approach to that
of~\cite{coecke:brunslakser}, which also gives an intuitionistic logic
for quantum mechanics.

The projections $\mathcal{P}(A)$ of any von Neumann algebra $A$ form
a complete orthomodular lattice~\cite{Redei}, and those in a Rickart C*-algebra form
a $\sg$-complete  orthomodular lattice \cite{Berberian}.\footnote{Orthomodularity is not mentioned in
\cite{Berberian}, but follows from the existence of a faithful representation.}
Recall that a lattice $\CL$ is called {\it orthomodular} when it is equipped with a
function $\perp \colon \CL \to \CL$ that satisfies:
\begin{enumerate}
 \item $x^{\perp\perp} = x$;
 \item $y^\perp \leq x^\perp$ when $x \leq y$;
 \item $x \wedge x^\perp = 0$ and $x \vee x^\perp = 1$;
 \item $x \vee (x^\perp \wedge y) = y$ when $x \leq y$.
\end{enumerate}
The first three requirements are sometimes called (1) ``double
negation'', (2) ``contraposition'', (3) ``noncontradiction'' and
``excluded middle'', but, as argued in Section~\ref{ss1.2}, one
should refrain from names suggesting a logical interpretation.
If these are satisfied, the lattice is called {\it orthocomplemented}. The
requirement (4), called the orthomodular law, is a weakening of
distributivity.

Any Boolean algebra is an orthomodular lattice, and any
orthomodular lattice is a combination of its Boolean sublattices, as
follows~\cite{KS,finch:structureofquantumlogic, kalmbach:orthomodularlattices}.
A \emph{partial Boolean algebra} is a family $(B_i)_{i \in I}$ of
Boolean algebras whose operations coincide on overlaps:
\begin{itemize}
 \item each $B_i$ has the same least element 0;
 \item $x \Raw_i y$ if and only if $x \Raw_j y$, when $x,y \in B_i \cap B_j$;
 \item if $x \Raw_i y$ and $y \Raw_j z$ then there is a $k \in I$ with
   $x \Raw_k z$;
 \item $\neg_i x = \neg_j x$ when $x \in B_i \cap B_j$;
 \item $x \vee_i y = x\vee_j y$ when $x,y \in B_i \cap B_j$;
 \item if $y \Raw_i \neg_i x$ for some $x,y \in B_i$, and $x \Raw_j z$
   and $y \Raw_k z$, then $x,y,z \in B_l$ for some $l \in I$.
\end{itemize}
These requirements imply that
\begin{equation}
\label{eq:partialBoolean}
 X = \bigcup_{i \in I} B_i
\end{equation}
carries a well-defined amalgamated structure $\vee, \wedge, 0, 1,
\perp$, under which it becomes an orthomodular lattice. For example,
$x^\perp = \neg_i x$ for $x \in B_i \subseteq X$.
Conversely, any orthomodular lattice $X$ is a partial Boolean algebra,
in which $I$ is the collection of all bases of $X$, and $B_i$ is the
sublattice of $X$ generated by $I$. Here, $B \subseteq X$ is called a
basis of $X$ when pairs $(x,y)$ of different
elements of $B$ are orthogonal, in the sense that $x \leq
y^\perp$. The generated sublattices $B_i$ are therefore automatically
Boolean. If we order $I$ by inclusion, then $B_i \subseteq B_j$ when
$i \leq j$. Thus there is an isomorphism between the categories of
orthomodular lattices and partial Boolean algebras.

A similar phenomenon occurs in the Heyting algebra defined by
\er{bohr2} when this is complete, which is the case for AW*-algebras
and in particular for von Neumann algebras (provided, of course, that we require $\CA$ to consist of commutative subalgebras
in the same class). Indeed, we can think of
$\mathcal{B}(\Sigma_A)$ as an amalgamation of Boolean
algebras: just as every $B_i$ in~\eqref{eq:partialBoolean} is a
Boolean algebra, every $\mathcal{P}(C)$ in~\eqref{bohr2} is a
Boolean algebra.
Hence the fact that the set $I$ in~\eqref{eq:partialBoolean} is
replaced by the partially ordered set $\mathcal{C}(A)$
in~\eqref{bohr2} and the requirement in~\eqref{bohr2} that $S$ be
monotone are responsible for making the partial Boolean algebra
$\mathcal{O}(\Sigma)$ into a Heyting
algebra (which by definition is distributive). Indeed, this
construction works more generally, as the
following proposition shows. Compare also~\cite{gravesselesnick:sheaves}.
\begin{proposition}
\label{prop:Heytingalgebra}
 Let $(I,\leq)$ be a partially ordered set, and $B_i$ an $I$-indexed
 family of complete Boolean algebras such that $B_i \subseteq B_j$ if
 $i \leq j$. Then
 \begin{equation}
 \label{eq:Heytingalgebra}
   Y = \{ f \colon I \to \bigcup_{i \in I} B_i \mid \forall_{i \in
     I}.f(i) \in B_i \mbox{ and } f \mbox{ monotone} \}
 \end{equation}
 is a complete Heyting algebra, with Heyting implication
 \begin{equation}
 \label{eq:Heytingimplication}
   (g \Raw h)(i) = \bigvee \{ x \in B_i \mid \forall_{j \geq i}.x \leq
     g(j)^\perp \vee h(j) \}.
 \end{equation}
\end{proposition}

It is remarkable that the lattice operations
on~\eqref{eq:Heytingalgebra} are defined pointwise, whereas the Heyting
implication~\eqref{eq:Heytingimplication} is not. But this ``nonlocality''  is necessary,
since a pointwise attempt $(g \Raw h)(i) = g(i) \Raw h(i)$ would not
provide a monotone function. We will also
write~\eqref{eq:Heytingimplication} as
\[
    (g \Raw h)(i)
  = \bigwedge_{j \geq i}^{B_i} g(j)^\perp \vee h(j),
\]
as in~\eqref{HI}.

\begin{proof}
 Defining operations pointwise makes $Y$ into a frame. For example,
 $(f \wedge g)(i) = f(i) \wedge_i g(i)$ is again a well-defined
 monotone function whose value at $i$ lies in $B_i$. Hence by a
 standard construction, $Y$ is a complete Heyting algebra by
 $g \Raw h = \bigvee \{ f \in Y \mid f \wedge g \leq h \}$. We now
 rewrite this Heyting implication to the form~\eqref{eq:Heytingimplication}:
 \begin{align*}
       (g \Raw h)(i)
   & = \big(\bigvee \{ f \in Y \mid f \wedge g \leq h \}\big)(i) \\
   & = \bigvee \{ f(i) \mid f \in Y, f \wedge g \leq h \} \\
   & = \bigvee \{ f(i) \mid f \in Y, \forall_{j \in I}. f(j) \wedge
       g(j) \leq h(j) \} \\
   & = \bigvee \{ f(i) \mid f \in Y, \forall_{j \in I} . f(j) \leq
       g(j)^\perp \vee h(j) \} \\
   & \stackrel{*}{=} \bigvee \{ x \in B_i \mid \forall_{j \geq i} . x \leq
       g(j)^\perp \vee h(j) \}.
 \end{align*}
  To finish the proof, we establish the marked equation.
  First, suppose that $f \in Y$ satisfies $f(j) \leq g(j)^\perp \vee
  h(j)$ for all $j \in I$. Take $x = f(i) \in B_i$. Then for all $j
  \geq i$ we have $x = f(i) \leq f(j) \leq g(j)^\perp \vee
  h(j)$. Hence the left-hand side of the marked equation is less than
  or equal to the right-hand side.
  Conversely, suppose that $x \in B_i$ satisfies $x \leq g(j)^\perp
  \vee h(j)$ for all $j \geq i$. Define $f \colon I \to \bigcup_{i \in
  I} B_i$ by $f(j) = x$ if $j \geq i$ and $f(j)=0$ otherwise. Then $f$
  is monotone and $f(i) \in B_i$ for all $i \in I$, whence $f \in Y$.
  Moreover, $f(j) \leq g(j)^\perp \vee h(j)$ for all $j \in I$. Since
  $f(i) \leq x$, the right-hand side is less than or equal to the
  left-hand side.
  \qed
\end{proof}

Hence every complete orthomodular lattice gives rise to a
complete Heyting algebra. The following proposition shows that the
former sits inside the latter.
\begin{proposition}
\label{prop:canonicalinjection}
 Let $(I,\leq)$ be a partially ordered set.
 Let $(B_i)_{i \in I}$ be a partial Boolean algebra, and suppose that
 every $B_i$ is  complete with $B_i \subseteq B_j$ for $i \leq j$. Then
 there is an injection $D \colon X \to Y$, where $X$ is the  complete
 orthomodular lattice as defined by~\eqref{eq:partialBoolean}, and $Y$
 is the corresponding   complete Heyting algebra as defined
 by~\eqref{eq:Heytingalgebra}. This injection reflects the order: if $D(x) \leq
 D(y)$ in $Y$, then $x \leq y$ in $X$.
\end{proposition}
\begin{proof}
 Define $D(x)(i) = x$ if $x \in B_i$ and $D(x)(i)=0$ if $x \not\in
 B_i$. Suppose that $D(x)=D(y)$. Then for all $i \in I$ we have $x
 \in B_i$ iff $y \in B_i$. Since $x \in X = \bigcup_{i \in I} B_i$,
 there is some $i \in I$ with $x \in B_i$. For that $i$, we have $x =
 D(x)(i) = D(y)(i) = y$. Hence $D$ is injective.

 If $D(x) \leq D(y)$ for $x,y \in X$, pick $i \in I$ such that $x \in
 B_i$. We have $x = D(x)(i) \leq D(y)(i) \leq y$. \qed
\end{proof}

The injection $D \colon X \to Y$ of the previous proposition is
canonical; for example, in the case of Theorem~\ref{maintheorem} the lattice
$Y=\mathcal{B}(\Sigma_A)$ is generated by the elements $D(x)$.
%\subsection{Implication}
We can use this to compare the logical structures of $X$ and $Y$.
Let us start with negation. The Heyting algebra $Y$ of
\eqref{eq:Heytingalgebra} has a negation $(\neg f) = (f \Raw 0)$. Explicitly:
\begin{equation}
\label{eq:Heytingnegation}
 (\neg f)(i) = \bigwedge_{j \geq i}^{B_i} f(j)^\perp.
\end{equation}
One then readily calculates:
\[
 D(x^\perp)(i) = \left[ \begin{array}{ll} 0 & \mbox{ if }x
     \not\in B_i \\ x^\perp & \mbox{ if }x \in B_i
   \end{array} \right],
 \quad
   (\neg (D(x)))(i)
 = \bigwedge_{j \geq i}^{B_i} \left[\begin{array}{ll}
     1 & \mbox{ if }x \not\in B_j \\
     x^\perp & \mbox{ if }x \in B_j
   \end{array}\right].
\]
For $x \not\in B_j$ and any $j \geq i$, we have
$D(x^\perp)(i)=0 \neq 1 = (\neg (D(x)))(i)$.
This situation already occurs for  $A=M_n(\C)$ with
$I=\CA$ and $X=\mathcal{P}(A)$.
Hence $D$ does not preserve negation.

We now turn to implication.
The Heyting algebra $Y$ of course has a Heyting implication $\Raw$
satisfying $f \wedge g \leq h$ iff $f \leq g \Raw h$.
The orthomodular lattice $X$ cannot have an implication, in general.
The best possible approximation of the Heyting implication $\Raw$ is
the \emph{Sasaki hook}
$\Raw_S$~\cite{dallachiaragiuntini:quantumlogics}, already given in \er{hook}.
This operation  satisfies the adjunction $x \leq y \Raw_S z$  iff  $x \wedge y \leq z$  only for $y$ and
$z$ that are compatible, in the sense that $y = (y \wedge z^\perp)
\vee (y \wedge z)$.
In fact, $y$ and $z$ are compatible if and only if they generate
a Boolean subalgebra, if and only if $y,z \in B_i$ for some $i \in I$.
In that case, the Sasaki hook $\Raw_S$ coincides with the implication
$\Raw_i$ of $B_i$.
Hence we find that
\begin{align*}
    (D(x) \Raw D(y))(i)
  & = \bigvee \{ z \in B_i \mid \forall_{j \geq i} . z \leq D(x)(j)
  \Raw_j D(y)(j) \}  \\
  & = \bigvee \{ z \in B_i \mid z \leq x \Raw_i y \} \\
  & = (x \Raw_S y).
\end{align*}
Thus the Sasaki hook $x \Raw_S y$ coincides with the Heyting
implication $D(x) \Rightarrow D(y)$ defined
by~\eqref{eq:Heytingimplication} at $i$ if $x$ and $y$ are compatible.
In particular, we find that $\Rightarrow$
and $\Raw_S$ coincide on $B_i \times B_i$ for $i \in I$; furthermore,
this is precisely the case in which the Sasaki hook satisfies the
defining adjunction for implications.
However, the canonical injection $D$ need not turn
Sasaki hooks into implications in general. One finds:
\begin{align*}
     D(x \Raw_S y)(i)
 & = \left[\begin{array}{ll}
       0 & \mbox{ if } x \not\in B_i \\
       x^\perp & \mbox{ if }x \in B_i, y \not\in B_i\\
       x^\perp \vee (x \wedge y) & \mbox{ if }x,y \in B_i
     \end{array}\right],\\
     (D(x) \Raw D(y))(i)
 & = \bigwedge_{j \geq i}^{B_i}
     \left[\begin{array}{ll}
       1 & \mbox{ if } x \not\in B_j \\
       x^\perp & \mbox{ if } x \in B_j, y \not\in B_j \\
       x^\perp \vee y & \mbox{ if } x,y \in B_j
     \end{array}\right].
\end{align*}
So for $x \not\in B_j$ and each $j \geq i$, we have
$D(x \Raw_S y)(i) = 0 \neq 1 = (D(x) \Raw D(y))(i)$.

Thus the canonical injection $D$ does not preserve negation in
general, nor does it turn Sasaki hooks into implications in
general. This shows that our intuitionistic quantum logic
\eqref{eq:Heytingalgebra} is of a very different nature than the
traditional quantum logic \eqref{eq:partialBoolean}, and argues in
favour of the Heyting implication~\eqref{eq:Heytingimplication}.

Another approach to intuitionistic quantum logic is to start with a
complete lattice
and perform the Bruns--Lakser completion~\cite{brunslakser,
coecke:brunslakser, stubbe:brunslakser}. The result is a complete
Heyting algebra which contains the original lattice join-densely, in
such a way that distributive joins that already exist are
preserved. Explicitly, the Bruns--Lakser completion of a lattice $L$ is the
collection $\mathrm{DI}(L)$ of its distributive ideals, ordered by
inclusion. Here, an ideal (lower set) $M$ is called distributive when
($\bigvee M$ exists and) $(\bigvee M) \wedge l = \bigvee_{m \in M} (m
\wedge l)$ for all $l \in L$.
We will now compare this Heyting algebra
with the one resulting from Proposition~\ref{prop:Heytingalgebra}, on
the example given by the orthomodular lattice $X$ that has the
following Hasse diagram.
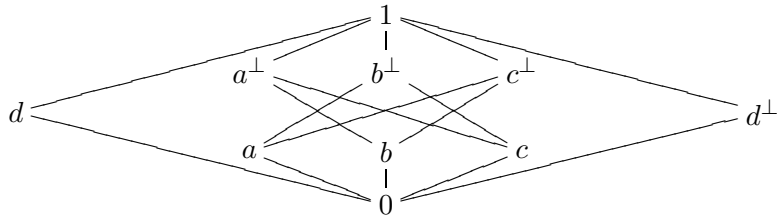
\[\xymatrix@C+2ex@R-5ex{
 &&& 1 \ar@{-}[ddl] \ar@{-}[dd] \ar@{-}[ddr] \ar@{-}[dddrrr]
 \ar@{-}[dddlll] \\ \\
 && a^\perp & b^\perp & c^\perp \\
 d & & & &&& d^\perp \\
 && a \ar@{-}[uur] \ar@{-}[uurr]
  & b \ar@{-}[uul] \ar@{-}[uur]
  & c \ar@{-}[uul] \ar@{-}[uull] \\ \\
 &&& 0 \ar@{-}[uul] \ar@{-}[uu] \ar@{-}[uur] \ar@{-}[uuulll]
 \ar@{-}[uuurrr]
}\]
This orthomodular lattice $X$ contains precisely five Boolean
algebras, namely $B_0=\{0,1\}$ and $B_i=\{0,1,i,i^\perp\}$ for $i \in
\{a,b,c,d\}$. Hence we take $I=\{0,a,b,c,d\}$ in
\eqref{eq:partialBoolean}, ordered by $i \leq j$ iff $B_i \subseteq
B_j$. Hence $i \leq j$ and $i \neq j$ imply $i=0$, and the monotony
requirement $\forall_{i \leq j}.f(i) \leq f(j)$ in
\eqref{eq:Heytingalgebra} becomes $\forall_{i \in \{a,b,c,d\}}.f(0)
\leq f(i)$. If $f(0) = 0 \in B_0$, this requirement is vacuous. But if
$f(0)=1 \in B_0$, the other values of $f$ are already fixed. Thus one
finds
\[
 Y \cong (B_1 \times B_2 \times B_3 \times B_4) + \{1\},
\]
which has 257 elements.

On the other hand, the distributive ideals of $X$ are given by
\begin{align*}
 \mathrm{DI}(X) & = \Big\{ \big(\bigcup_{x \in A} \downset x\big) \cup
                  \big(\bigcup_{y \in B} \downset y\big) \;\Big|\;
                  A \subseteq \{a,b,c,d,d^\perp\},
                  B \subseteq \{a^\perp,b^\perp, c^\perp\}\Big\}  \\
       & - \{\emptyset\} + \{X\}.
\end{align*}
In the terminology of~\cite{stubbe:brunslakser},
\[
 \mathcal{J}_{\mathrm{dis}}(x) = \{ S \subseteq \downset x \mid x \in S\},
\]
\textit{i.e.}~the covering relation is the trivial one, and
$\mathrm{DI}(X)$ is the Alexandrov topology (as a frame/locale).
We are unaware of instances of the 
Bruns--Lakser completion of orthomodular lattices that occur naturally
in quantum physics but lead to Heyting algebras different from ideal
completions.
The set $\mathrm{DI}(X)$ has 72 elements. 

The canonical injection $D$
of Proposition~\ref{prop:canonicalinjection} 
need not preserve the order, and hence does not satisfy the universal
requirement of which the Bruns--Lakser completion is the solution.
Therefore, it is unproblemetic to conclude that the construction in
Proposition~\ref{prop:Heytingalgebra} differs from the Bruns--Lakser
completion. 

\section{Measures on projections and pairing formula}\label{finalsection}
Theorem 14 in \cite{HLS} gives a bijective correspondence between quasi-states on a C*-algebra $A$
and internal probability valuations on the Gelfand spectrum $\CO(\Sg_{\uA})$. In case that
 $A$ is  a Rickart C*-algebra, we can say a bit more.
We start by recalling a few definitions, in which $[0,1]_l$ is the collection of lower reals between
0 and 1, and $[0,1]$ denotes the Dedekind reals.
\begin{definition}
\begin{enumerate}
\item
A \emph{probability  measure} on a $\sg$-complete
 orthomodular lattice  $\CL$ is a function $\mu:\CL\raw [0,1]$
 that  on any $\sg$-complete Boolean sublattice of $\CL$
 restricts to a probability measure (in the traditional sense). 
 \item
 A \emph{probability valuation} on a Boolean lattice $L$
is  a function $\mu:L\to [0,1]_l$ such that
\begin{enumerate}
\item $\mu(0)=0$, $\mu(1)=1$;
\item if $x\leq y$, then $\mu(x)\leq \mu(y)$;
\item $\mu(x)+\mu(y)=\mu(x\wed y)+\mu(x\vee y)$.
\end{enumerate}
\item
  A \emph{continuous probability valuation} on a compact regular frame $\CO(X)$ is a monotone function
  $\nu:\CO(X)\raw [0,1]_l$ that satisfies $\nu(1)=1$ as well as  $\nu(U)+\nu(V)=\nu(U\wed
  V)+\nu(U\vee V)$ and $\nu(\bigvee_{\lm} U_{\lm}) =\bigvee_{\lm}\nu(
  U_{\lm})$ for every directed family.
  \end{enumerate}
\end{definition}
 We will apply part 1 of this definition to $\CL=\CP(A)$ in \Sets; see Proposition \ref{eigRCA}
 for its $\sigma$-completeness.
 Part 2 will be applied internally  to $L=\underline{\CP}_{\underline{A}}$
in $\TA$ (i.e.\ the functor $C\mapsto\CP(C)$).
As to part 3,
if $X$ is a compact Hausdorff space in \Sets, a continuous probability valuation on $\CO(X)$ is essentially the same thing as a regular probability  measure on $X$. We will actually apply the definition internally  to the frame $\CO(\uSg_{\uA})$ in $\TA$.
\begin{theorem}\label{last}
Let $A$ be a  Rickart C*-algebra.
There is a bijective correspondence between:
\begin{enumerate}
\item quasi-states on $A$;
\item
 probability  measures on  $\mathcal{P}(A)$;
 \item
  probability valuations on the Boolean lattice  $\underline{\CP}_{\underline{A}}$ in $\CT(A)$;
 \item
continuous probability  valuations on the Gelfand spectrum $\CO(\underline{\Sg}_{\underline{A}})$ in $\CT(A)$.
\end{enumerate}
\end{theorem}
\begin{proof}
We include the first item only for completeness; the equivalence between
1 and 4 is contained in Theorem 14 in \cite{HLS}.
The equivalence between 3 and 4 follows from Theorem \ref{RCAS} and
the observation in~\cite[\S 3.3]{integrals-valuations} that valuations on a
compact regular frame are determined by their behaviour on a generating lattice;
indeed, if a frame $\CO(X)$ is generated by $L$, then a probability measure $\mu$
on $L$ yields a continuous probability  valuation $\nu$ on $\CO(X)$ by
$\nu(U)=\sup \{\mu(u) \mid u \in U\}$, where $U\subset L$ is regarded as an element of $\CO(X)$.

To prove the equivalence between 2 and 3 we use the following lemma,  which
holds in the internal logic of any topos\footnote{Classically, this lemma is trivial as the lower
reals and the Dedekind reals coincide.}.
\begin{lemma}
Let $L$ be a Boolean algebra and $\mu$ a valuation on $L$. Then
$\mu(x)$ is a Dedekind real for every $x\in L$.
\end{lemma}
\begin{proof}
Let $s+\epsilon<t$ in $\Q$. We need to prove that $s<\mu(x)$ or $\mu(x)\leq t$.
The last statement is defined as $\mu(y)>1-t$ for some $y$ such that
$x\wedge y=0$. We choose $y=x^\perp$, the complement of $x$. Now,\[
1-\epsilon < \mu(\top) \text{ and } s+\epsilon-t<0=\mu(\bot),
                          \]
equivalently,
\[1-\epsilon \leq \mu(x\vee x^\perp) \text{ and } s+\epsilon-t<\mu(x\wedge
x^\perp).\]
By the modular law for valuations and Lemma 2.2 in \cite{integrals-valuations},
if $p+q<\mu(z\wedge w)+\mu(z\vee w)$, then
$p<\mu(w)$ or $q<\mu(z)$. Choosing $z=x, w=x^\perp, p=s ,q=1-t$ we have
\[ s<\mu(x) \text{ or } 1-t<\mu(x^\perp).\]
That is,
\[ s<\mu(x) \text{ or } \mu(x)\leq t.\]
It follows that $\mu(x)$ is a Dedekind real.\qed \end{proof}

Since the Dedekind reals  in $\TA$ are internalized as  the constant functor $\underline{\R}:C\mapsto\R$ (as opposed to the lower reals),
according to this lemma
an internal probability valuation $\nu: \underline{\CP}_{\underline{B(H)}}\raw\underline{[0,1]}$ is defined by its components
(as a natural transformation) $\nu_C: \CP(C)\raw [0,1]$. By naturality, for $p\in \CP(C)$, the number
$\nu_C(p)\equiv \mu(p)$ is independent of $C$, from which the equivalence between
2 and 3 in Theorem \ref{last} is immediate.
\qed\end{proof}

Finally, we justify the formula \er{result} in case $A=B(H)$ for some
Hilbert space $H$, by identifying $V_{\ps}(S)$ with the nonprobabilistic state-proposition
pairing $\langle S,\psi\rangle$ defined in \cite{HLS}; see Section 6 of that paper for the background of the following computation. By definition, $C\in \langle S,\psi\rangle$ iff  $C\Vdash \nu^{\psi}(S)=1$, where $\nu^{\psi}$ is the probability valuation
on $\CO(\underline{\Sg}_{\underline{B(H)}})$ defined by a normal state $\psi$ on $B(H)$, seen as a probability  measure on  $\mathcal{P}(B(H))$. Using \er{combi}, we describe $S\in \CO(\underline{\Sg}_{\underline{B(H)}})$ as a subfunctor $\underline{U}$
of $\underline{\CP}_{\uA}$, which (lying in the set of ideals) is locally closed under
$\vee$.  Then the  following are equivalent:
\begin{eqnarray*}
&C\Vdash& \nu^{\psi}(\underline{U})=1\\
&C\Vdash& \forall q<1. \nu^{\psi}(\underline{U})>q\\
\text{for all }D\supseteq C \text{ and } q<1, &D\Vdash&
\nu^{\psi}(\underline{U})>q\\
\text{for all }D\supseteq C \text{ and } q<1, &D\Vdash& \exists u \in
\underline{U}. \nu^{\psi}(u)>q\\
\text{for all }D\supseteq C \text{ and } q<1, \text{ there exists } u\in
\underline{U}(D) \text{ s.t.}
&D\Vdash& \nu^{\psi}(u)>q\\
\text{for all }q<1, \text{ there exists } u\in \underline{U}(C) \text{ s.t.\ }
\nu^\psi(u)>q\\
\sup_{u\in \underline{U}(C)}\nu^\psi(u)=1\\
\nu^\psi(\underline{U}(C))=1.
\end{eqnarray*}
Now $\underline{U}(C)$ is a collection of projections. By classical
meta-logic we can take its supremum $p:=\bigvee \underline{U}(C)$. Then
$\psi(p)=\nu^\psi(p)=1$, which proves \er{result}.
\bibliographystyle{plain}
\bibliography{Synthesev4,bibliography}
\end{document}